\documentclass[review]{elsarticle}

\usepackage{graphicx} 
\usepackage{amsmath} 
\usepackage{lineno,hyperref}
\usepackage{setspace}
\usepackage{amssymb}
\usepackage{color}
\usepackage{adjustbox}

\begin{document}

\begin{frontmatter}
	
\title{New and Extended Data Processing of Mars Odyssey Neutron Spectrometer Data}
	
\author[lanl]{K.E. Mesick\corref{cor1}}\ead{kmesick@lanl.gov}
\author[psi]{W.C. Feldman}
\author[lanl]{E.R. Mullin}
\author[lanl]{L.C. Stonehill}
	
\address[lanl]{Los Alamos National Laboratory, Los Alamos, NM 87545 USA}
\address[psi]{Planetary Science Institute, Tuscon, AZ 85719 USA}
	
\cortext[cor1]{Corresponding author}
\begin{abstract}
The Los Alamos National Laboratory designed and built Mars Odyssey Neutron Spectrometer (MONS) has been operating and collecting data from February 2002 to the present.  MONS measures the neutron leakage albedo from galactic cosmic ray bombardment of Mars.  These signals can indicate the presence of near-surface water deposits on Mars, and can also be used to study properties of the seasonal polar CO$_2$ ice caps.  This work outlines a new analysis of the MONS data that results in new and extended time-series maps of MONS thermal and epithermal neutron data.  The new data are compared to previous publications on the MONS instrument.  We then present preliminary results studying the inter-annual variability in the polar regions of Mars based on 8 Mars-Years of MONS data from the new dataset. 
\end{abstract}
	
	
\end{frontmatter}


\section{Introduction}

The Los Alamos National Laboratory (LANL) designed and built Mars Odyssey Neutron Spectrometer (MONS) has been continuously measuring the leakage flux of neutrons from a Mars polar orbit since February 2002.   These data have been used to map the hydrogen content of Mars up through July 2009 \cite{Maurice2011,Feldman2011}.  However, both the Odyssey spacecraft and the MONS instrument have been operational to the present time; here we present a new and extended analysis of the MONS dataset through December 2017 that can be used to search for long-term climate variations, particularly in the polar regions of Mars.  Processing of the integrated total dataset is an important step before such interpretations can be drawn.   Throughout this process, many choices and alternatives must be clearly documented to establish the accuracy, precision, and robustness of these data.

The MONS instrument collects neutron fluxes continuously from the Mars surface in three energy bands: thermal (0--0.4 eV), epithermal (0.4 eV--700 keV), and fast (0.7--5 MeV). The purpose of this work is to transform time-tagged measurements through the present time into relevant neutron maps, which are time dependent because of seasonally changing CO$_2$- and water-ice precipitation at high latitudes.  MONS has continued to collect data over 17 years of operation to present, well over its initial science phase duration. However, over that time period, there have been several unresolved issues regarding our understanding of the MONS systematic biases.  At present, most, if not all, of these biases have been removed. Each generation of MONS data processing was built independently of its predecessors, often by different people, to limit the propagation of erroneous assumptions. Efforts were also devoted to compare the results of each approach. 

The initial processing of MONS data was performed by Tokar \textit{et al.} \cite{Tokar2002} and was used for early discovery results. Subsequently, Prettyman \textit{et al.} developed an independent approach \cite{Prettyman2004} that has been the reference for publications between 2004 to present. This code is currently used to deliver level‐1 derived neutron data (DND) to the Planetary Data System (PDS). These products are time series of corrected neutron counting rate data that can be used for scientific investigations. The level‐1 dataset includes averaged neutron data (AND), which consists of neutron maps built from the neutron time series data. The most recent processing and analysis of MONS data was performed by Maurice \textit{et al.} \cite{Maurice2011} and covered data through July 2009.  This led to work studying the depth-distribution of water on Mars \cite{Feldman2011,Pathare2018}.  

On the way to any science interpretation of inter-annual variability in the MONS dataset, this paper intends to document and provide the necessary elements for understanding the new data processing method and resulting dataset.  We then build on this stage by presenting averaged counting rate maps and a preliminary comparison of the inter-annual variability in the Mars polar regions over 8 Mars Years by means of the neutron counting rates.  These can be compared to previous results presented in \cite{Prettyman2004,Prettyman2009,Maurice2011}.

\section{MONS Instrument}

\begin{figure}[h]
\centering
\includegraphics[width=0.8\textwidth]{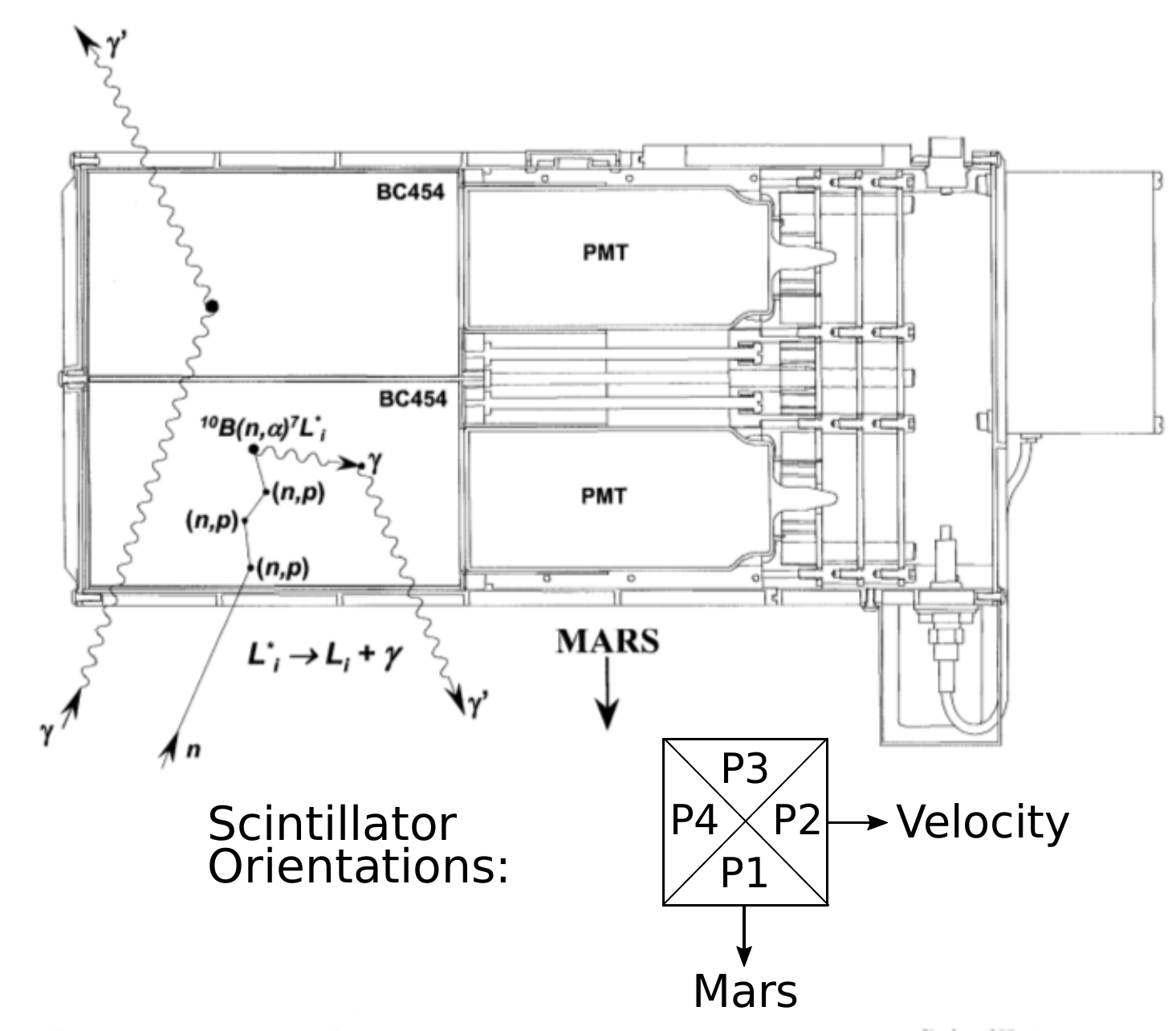}
\caption{Schematic of the MONS instrument and cartoon of scintillator orientation.}
\label{fig:mons}
\end{figure}

The MONS instrument consists of an 11$\times$11$\times$10~cm$^3$ BC454 plastic scintillator separated into four optically isolated segments, or ``prisms." This plastic is loaded with 5\% natural boron by weight, which provides sensitivity to thermal and epithermal neutrons through neutron capture on $^{10}$B.  The predominant interaction that occurs is $^{10}$B(n,$\alpha$)$^{7}$Li$^{*}$ with a Q value of 2.8 MeV.  Due to inefficiency in light production from the heavy isotopes produced in this reaction, this energy is quenched and detected at 98~keV electron equivalent (keVee).  A schematic of the MONS instrument is shown in Fig.~\ref{fig:mons}.

There are two types of primary data products produced by the MONS instrument that define what type of neutron was detected \cite{Maurice2011}.  Category 1 events (thermal, epithermal) are defined by a prompt interaction with an energy between 40~keVee and 630~keVee that is not followed by a delayed interaction within 25.6~$\mu$s.  Category 2 events (fast) are defined as a similar prompt pulse with an expanded energy range of 40~keVee to 2.55~MeVee followed by a delayed pulse with an energy between 40~keVee and 630~keVee within a 25.6~$\mu$s window.  For both Category 1 and Category 2 events, only one or two prisms can detect the event, otherwise the event is thrown away. Prompt events with an energy greater than 2.55~MeVee are categorized as GCR events.

Fast neutrons (Category 2 events) are defined as neutrons with an energy $>$0.7~MeV \cite{Maurice2011} and can be detected by all four prisms.  Category 1 events can be split into thermal and epithermal neutrons based on prism.  Prism 1 faces the nadir direction and is covered with a 0.69~mm thick cadmium sheet, which absorbs neutrons below $\sim$0.4 eV.  Therefore, Category 1 events from this prism are epithermal neutrons (0.4~eV - 0.7~MeV).  As noted in \cite{Maurice2011,Prettyman2009}, due to the geometry of the prisms there are small gaps in the cadmium coverage allowing Prism 1 some thermal neutron sensitivity.  Thermal neutrons are detected by exploiting the Doppler filter technique \cite{Feldman1986}, which uses the fact that the spacecraft velocity (3.4~km/s) is faster than the velocity of thermal neutrons (most probable value of 1.9~km/s or 0.019 eV at Mars' atmosphere temperature).  Prism 2 is forward facing along the direction of spacecraft motion, and therefore detects both thermal and epithermal neutrons.  Prism 4 faces backwards along the direction of spacecraft motion, and therefore only neutrons that have a velocity higher than the spacecraft velocity can be detected.  This corresponds to neutrons in the epithermal range with an energy greater than 0.06 eV.  Thermal neutron ($<$0.06~eV) rates are determined by subtracting the Prism 4 counting rate from the Prism 2 counting rate.  Similarly, an alternate definition of epithermal neutrons (0.06~eV - 0.7 MeV) can be obtained from the Prism 4 counting rates.  Finally, Prism 3 is shielded from Mars and therefore should be a good proxy for the spacecraft background.  The sides of the prisms are also covered in cadmium.

The mapping phase of the MONS instrument began February 22, 2002 and has been operating nearly continuously since then, leading to 17 Earth-years of data.  The most recent processing and analysis of MONS data \cite{Maurice2011} covered data through July 2009, corresponding to nearly 4 Mars-years of data ($L_s$ = 330 in Mars Year (MY) 25 to $L_s$ = 313 in MY 29, using the Mars calendar defined by \cite{Piqueux2015}). While \cite{Maurice2011} showed some inter-annual comparisons of counting rates in the polar regions, their work focused primarily on creating an averaged CO$_2$ frost-free map of two-layer water-equivalent hydrogen (WEH) based on the MONS data that subsequently was used in the most definitive MONS mapping of WEH and its depth distribution to date \cite{Pathare2018}.  Another paper including inter-annual comparisons of the CO$_2$ frost cap thickness for two Mars years towards the beginning of the MONS mission can be found in \cite{Prettyman2009}.  Here we present new data processing of the Category 1 MONS data that includes all data through the end of 2017 ($L_s$ = 108.3 in MY 34).  This doubles the amount of data processed by \cite{Maurice2011} and quadruples the number of MY in a detailed comparison of inter-annual variability of the seasonal CO$_2$ frost deposits in the polar regions.

\section{New Data Processing}

The data processing includes many steps to take the MONS data from raw binary data to prism counting rates registered with latitude and longitude.  Much of the data processing follows and draws upon the work described in \cite{Maurice2011}, but performed independently.  Raw data were acquired from the Planetary Data System (PDS) Geosciences Node (\url{pds-geosciences.pds.wustl.edu}), which releases data quarterly for the Mars Odyssey mission and GRS instrument suite.  The raw data or experimental data records (EDR) are organized into folders by calendar year and subsequently by day.  Raw data for the MONS instrument are contained within the neutron\_spectra files.  Relevant engineering data for MONS are contained within the eng subdirectory.  Information on the format of each binary EDR file is contained within the main label directory.

The MONS data are pre-packaged to contain ephemeris data in addition to the instrument data.  Each data point is registered with a UTC time stamp and an ``SLCK" clock value that is unique for each data point.  The neutron data includes 64-channel histograms for Category 1 events and 32-channel histograms for the prompt (early) and delayed (late) Category 2 events.  Counter data, which store the total number of counts over threshold in 19.75~second accumulation windows, include GCR, deadtime, and the number of, and which, prisms fired.  There is additional information on the first 84 Category 2 events within each accumulation window, including time between the prompt and delayed pulses, and pulse heights.  The data also contain sub-satellite latitude and longitude at the middle of each integration window and position and velocity of the spacecraft in different reference frames.

The raw data conversion was done using Python 3.5 and the unpacked data stored in a MySQL Database.  Following conversion, data reduction takes place to remove bad data from the dataset, described in Section~\ref{sec:data_reduction}.  After all bad data are removed, data corrections that result in the final dataset are applied, described in Section~\ref{sec:data_correction}.

\subsection{Data Reduction}\label{sec:data_reduction}

There are several categories of ``bad" data that must be removed before further processing can take place.  These data cuts apply to both Category 1 and Category 2 MONS data, although only Category 1 data are processed here.  The first and largest data cut is from solar energetic particle (SEP) events, that produce a large background in the prism counting rates.  Stability cuts are also applied to the counter data, which remove outliers and transients in these datasets.  Cuts on spacecraft orbit parameters are applied to also remove outliers or transients and remove data acquired during clock resets that corrupt our ability to normalize to counting rates.  Finally, some additional data cuts related to various anomalous readings are applied.  The final dataset contains only good data that passes all four of the following described cuts.  A summary of how much data is removed by each cut is provided at the end of this sub-section.

\subsubsection{SEP Event Cuts}

\begin{figure}[h]
\centering
\includegraphics[width=0.8\textwidth]{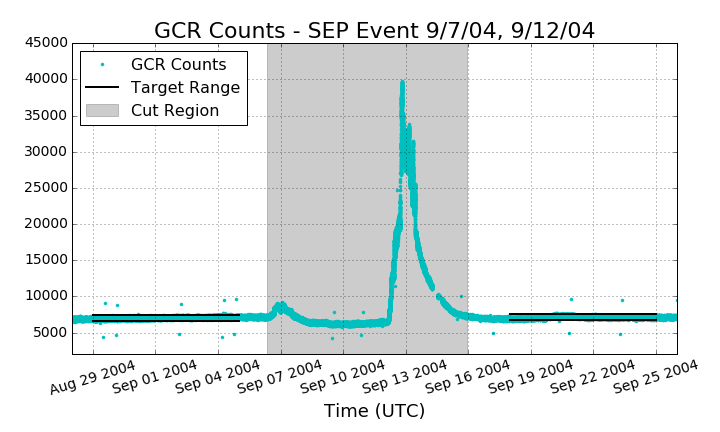}
\caption{Example of default SEP event cut definition for September 2004 events.  See text for details.}
\label{fig:sep1}
\end{figure}

Removal of SEP events is done manually by looking at the counting rate recorded by a dedicated GCR counter and removing periods of rate excursions.  An example of the base procedure is described below for a SEP event in September 2004, shown in Fig.~\ref{fig:sep1}.  The mean and standard deviation of the GCR counter (total counts in each 19.75 s accumulation window) is determined for 8 days before and 8 days after the event.  Figure~\ref{fig:sep1} shows black bands representing $\pm$3 standard deviations ($\sigma$) from the mean.  The excursion is flagged as when the GCR counter extends beyond $\pm$3$\sigma$ from these means.  To safely remove the full extent of each event, the SEP event cut range starts 4 hours before the start of the excursion and ends 4 hours after the end of the excursion.  The final cut range is demonstrated as the gray shaded region.  

\begin{figure}[h]
\centering
\includegraphics[width=0.8\textwidth]{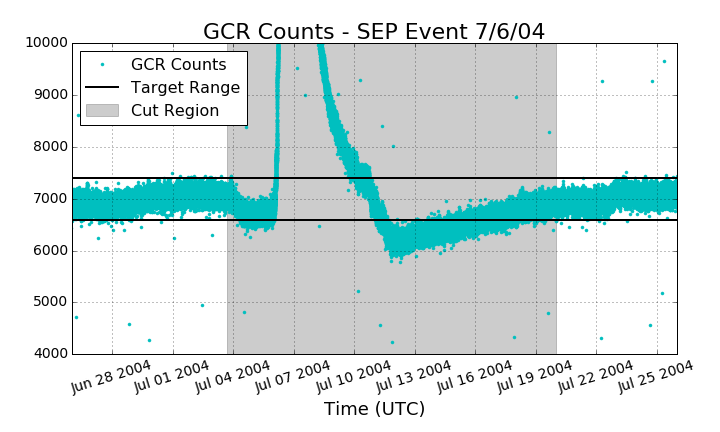}
\caption{Example of modified SEP event cut definition for July 2004 event.  See text for details.}
\label{fig:sep2}
\end{figure}

There are several SEP events where the event cut method was adapted or event cut ranges were manually updated from the base method.  These included SEP events that were low in strength or short in duration, but most frequently were when a decrease in the GCR counter was observed surrounding the peak of the SEP event (likely due to changes in the interplanetary magnetic field).  This was observed in $\sim$20\% of SEP events.  In these cases if the dip was before the main SEP excursion, the start of the event cut was determined by eye.  If the dip was after the main SEP excursion, the mean and standard deviation in counter from before the event was used to judge when the counter returned to nominal.  An example of this type of event (July 2004) is shown in Fig.~\ref{fig:sep2}.  This event exhibited the decrease in rates both before and after the event.  The mean was determined from data between 6/26/2004 -- 7/3/2004.

Ranges defining the removal of SEP events from the dataset are summarized in Table~\ref{table:sep_cut}.

\begin{table}[h!]
\caption{Event cut ranges for removing SEP events, in UTC range.}
\label{table:sep_cut}
\centering
\begin{adjustbox}{width=0.95\textwidth}
	\begin{tabular}{|l|c|c||l|c|c|}
		\hline
		Event & Start Time (UTC) & End Time (UTC) & Event & Start Time (UTC) & End Time (UTC) \\
		\hline
		1 & 2002-03-11 00:00 & 2002-03-20 10:00 & 36 & 2012-05-16 22:00 & 2012-05-21 14:00 \\
		2 & 2002-04-21 23:00 & 2002-05-05 10:00 & 37 & 2012-07-06 21:00 & 2012-08-04 10:00 \\
		3 & 2002-05-21 18:00 & 2002-05-29 00:00 & 38 & 2012-08-31 21:00& 2012-09-04 14:00 \\
		4 & 2002-07-15 20:00 & 2002-08-06 12:00 & 39 & 2012-09-20 00:00& 2012-10-01 18:00 \\
		5 & 2002-08-14 00:00 & 2002-09-18 06:00 & 40 & 2013-03-05 04:00& 2013-03-09 18:00 \\
		6 & 2002-10-14 11:00 & 2002-10-20 00:00 & 41 & 2013-05-01 04:00& 2013-05-02 14:00 \\
		7 & 2002-10-24 12:00 & 2002-11-14 00:00 & 42 & 2013-05-12 23:00& 2013-05-16 22:00 \\
		8 & 2002-12-02 12:00 & 2002-12-05 00:00 & 43 & 2013-05-23 12:00& 2013-05-29 00:00 \\
		9 & 2003-03-18 13:00 & 2003-03-28 16:00 & 44 & 2013-08-19 22:00& 2013-08-25 14:00 \\
		10 & 2003-05-28 13:00 & 2003-06-02 16:00 & 45 & 2013-10-05 05:00& 2013-10-09 14:00 \\
		11 & 2003-10-25 06:00 & 2003-11-25 04:00 & 46 & 2013-10-11 03:00& 2013-10-18 10:00 \\
		12 & 2003-12-02 13:00 & 2003-12-05 12:00 & 47 & 2013-11-02 03:00& 2013-11-15 14:00 \\
		13 & 2004-07-03 16:00 & 2004-07-20 00:00 & 48 & 2013-12-26 05:00& 2014-01-01 12:00 \\
		14 & 2004-09-06 08:00 & 2004-09-15 22:00 & 49 & 2014-01-06 03:00& 2014-01-14 06:00 \\
		15 & 2004-11-10 23:00 & 2004-11-19 20:00 & 50 & 2014-02-14 12:00& 2014-03-18 12:00 \\
		16 & 2005-01-11 12:00 & 2005-02-04 14:00 & 51 & 2014-03-29 18:00& 2014-03-31 00:00 \\
		17 & 2005-05-14 00:00 & 2005-05-20 00:00 & 52 & 2014-04-18 15:00& 2014-04-22 18:00 \\
		18 & 2005-06-16 17:00 & 2005-06-24 10:00 & 53 & 2014-05-09 03:00& 2014-05-12 00:00 \\
		19 & 2005-07-14 05:00 & 2005-08-09 22:00 & 54 & 2014-09-01 10:00& 2014-09-15 14:00 \\
		20 & 2005-08-22 14:00 & 2005-09-23 13:00 & 55 & 2014-09-22 00:00& 2014-10-01 00:00 \\
		21 & 2006-11-03 18:00 & 2005-11-10 03:00 & 56 & 2014-10-15 00:00& 2014-10-20 00:00 \\
		22 & 2006-12-05 07:00 & 2006-12-20 18:00 & 57 & 2014-11-01 00:00& 2014-11-03 00:00 \\
		23 & 2007-01-25 05:00 & 2007-01-27 05:00 & 58 & 2014-11-07 00:00& 2014-11-11 00:00  \\
		24 & 2010-06-11 22:00& 2010-06-13 04:00& 59 & 2014-12-13 00:00& 2014-12-28 13:00 \\
		25 & 2010-08-05 06:00& 2010-08-09 10:00& 60 & 2015-03-03 12:00& 2015-03-12 00:00 \\
		26 & 2011-02-11 12:00& 2011-02-12 12:00& 61 & 2015-03-23 22:00& 2015-04-01 20:00 \\
		27 & 2011-03-08 00:00& 2011-04-11 18:00& 62 & 2015-04-21 11:00& 2015-04-24 09:00 \\
		28 & 2011-05-09 20:00& 2001-05-11 22:00& 63 & 2015-05-02 10:00& 2015-05-08 12:00 \\
		29 & 2011-06-04 19:00& 2011-06-12 13:00& 64 & 2015-06-18 00:00& 2016-06-24 00:00 \\
		30 & 2011-07-26 00:00& 2011-07-29 00:00& 65 & 2015-10-28 13:00& 2015-11-05 00:00 \\
		31 & 2011-09-04 00:00& 2011-10-08 14:00& 66 & 2016-01-06 00:00& 2016-01-07 15:00 \\
		32 & 2011-11-03 19:00& 2011-11-08 23:00& 67 & 2016-02-21 12:00& 2016-02-23 12:00 \\
		33 & 2011-11-29 12:00& 2011-11-30 12:00& 68 & 2016-03-16 12:00& 2016-03-17 18:00 \\
		34 & 2012-01-23 06:00& 2012-02-05 04:00& 69 & 2017-04-14 18:00& 2016-04-21 00:00 \\
		35 & 2012-03-06 23:00 & 2012-03-17 10:00 & 70 & 2017-09-10 15:00 & 2017-09-21 12:00 \\
		\hline
	\end{tabular}
\end{adjustbox}
\end{table}

\subsubsection{Stability Cuts}

Stability cuts were applied to the GCR counters and the total counts in each of the four prism Category 1 histograms.  The stability cuts are applied based on the deviation of each data point from a boxcar rolling median value.  A rolling window is specified as the number of data entries to sum over, and the result is centered within the time range of the window.  Based on the time scales over which observed rates can change, we chose to apply a ``daily" rolling median window.  Resampling data from 2003 through 2007 to a frequency of one day, the typical number of data entries in one day was 4185 entries.

\begin{figure}[h]
\centering
\includegraphics[width=0.42\textwidth]{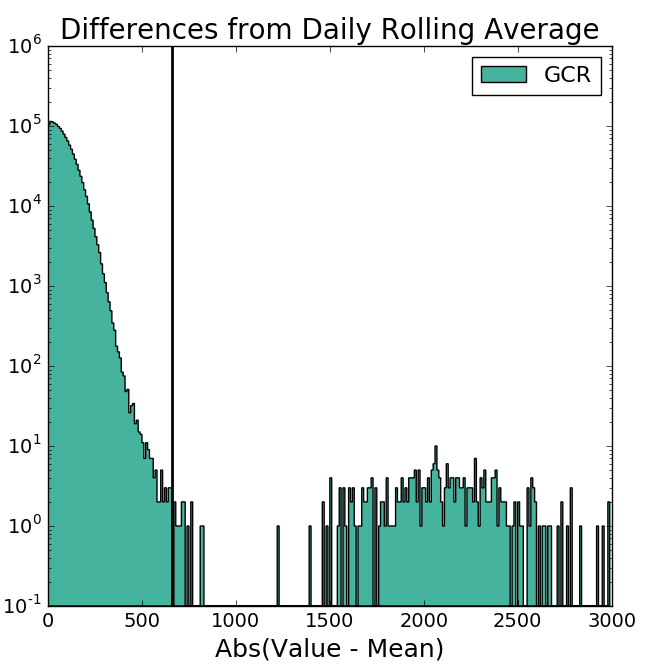}
\hspace{0.2 in}
\includegraphics[width=0.42\textwidth]{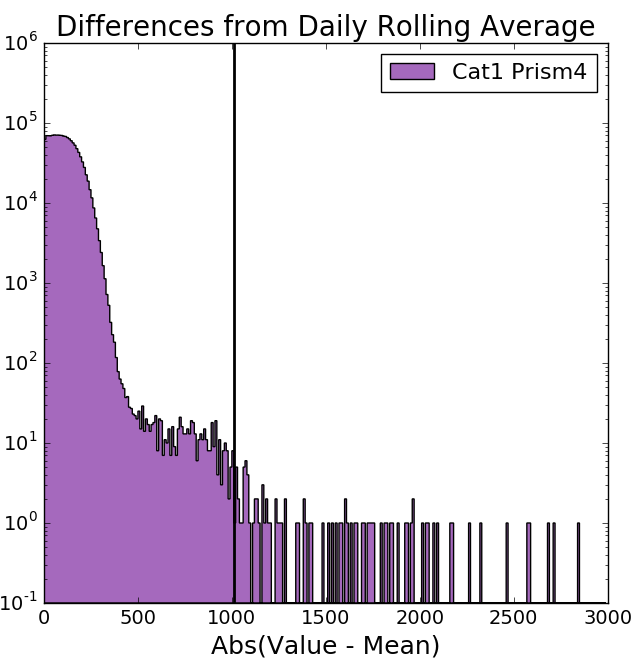}
\caption{Histogram of the deviations from the rolling median with the determined stability threshold for the GCR counter (left) and Category 1 Prism 4 counts (right).}
\label{fig:stab1}
\end{figure}

\begin{figure}[h!]
\centering
\includegraphics[width=0.75\textwidth]{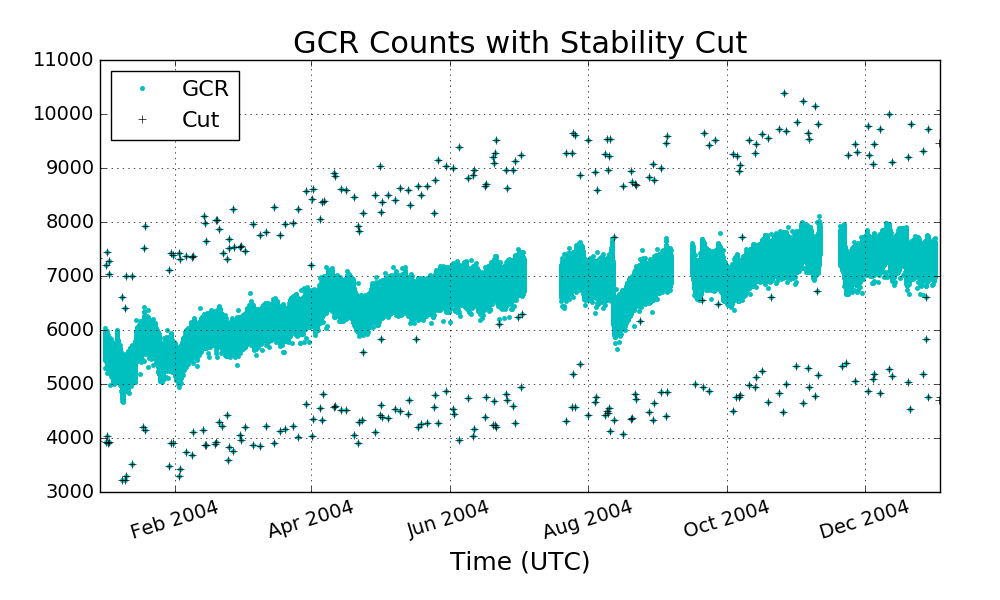}
\includegraphics[width=0.75\textwidth]{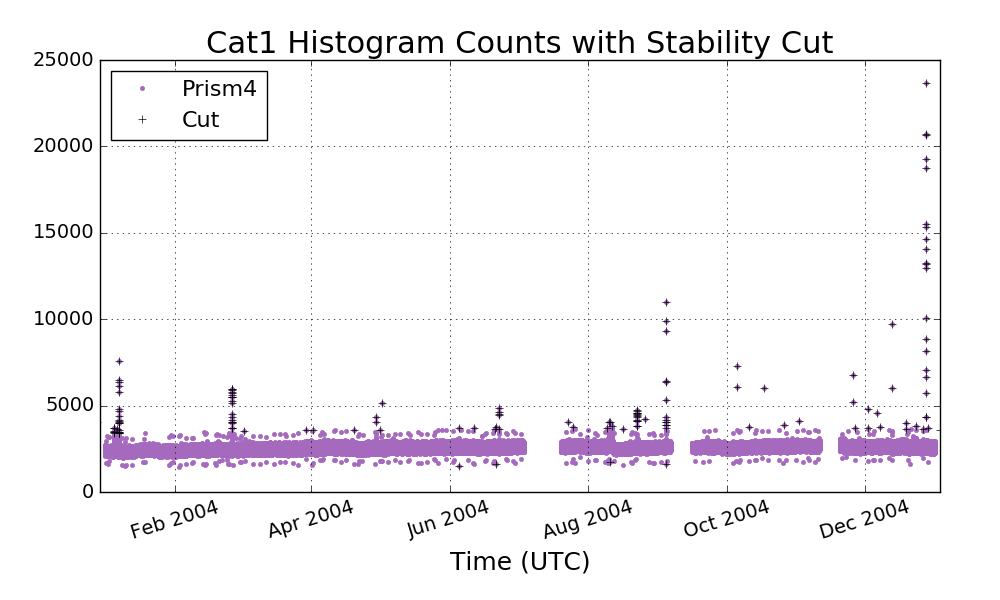}
\caption{Example of the stability cut applied.  See text for details.}
\label{fig:stab2}
\end{figure}

An example of this technique is described using data from 2004.  Figure~\ref{fig:stab1} shows a histogram of the deviations from the rolling median value for the GCR counter (left) and the Category 1 Prism 4 total histogram counts (right).  The line indicates the stability threshold, which was chosen as ten times the median of the deviations (sometimes called the ``MAD").  Data greater than the stability threshold are cut, indicated by the + markers in Fig.~\ref{fig:stab2}.  The stability cut most obviously affects what seem to be spurious readings in the GCR counter.  For the Prism total histogram counts the stability cuts remove most spikes observed in the rates.


\subsubsection{Spacecraft Cuts}

\begin{figure}[h]
\centering
\includegraphics[width=0.65\textwidth]{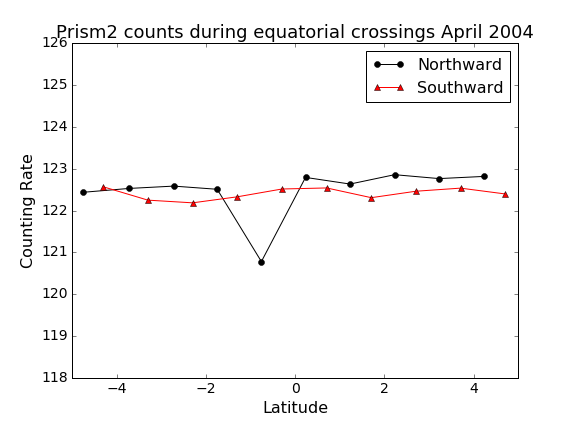}
\caption{(Color online) Prism 4 count rates assuming the nominal 19.75~s window in April 2004 when crossing the equator Northward (black circles) and Southward (red triangles).}
\label{fig:eqcross}
\end{figure}

Cuts were applied based on the spacecraft orbit data to flag when a variable goes out of a range or experiences an excursion from nominal.  First, spacecraft orientation data were required to be available in the raw data (flags for pointing and intersecting equal to 1).  The parameters subject to spacecraft cuts that must be between certain values are the latitude ($-90^{\circ}$ to $+90^{\circ}$), longitude ($0^{\circ}$ to $360^{\circ}$), and altitude (380~km to 460~km).  In addition, issues related to the Northerly equatorial crossing is flagged under this cut.  As described in \cite{Maurice2011} and illustrated in Fig.~\ref{fig:eqcross}, when the spacecraft is moving Northward and crosses the equator the internal clock is reset and thus the measurement time interval is lost.  These events cannot be properly normalized into counting rates and are removed.  The most common bounding cuts come from pointing and intersecting data not being available.


Finally, some instances of transient deviations in the mars position and velocity as recorded in the instrument frame were observed and removed.

\subsubsection{Other Cuts}

\begin{figure}[h!]
\centering
\includegraphics[width=0.45\textwidth]{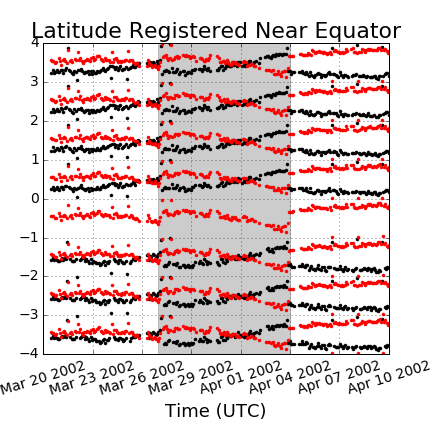}
\includegraphics[width=0.45\textwidth]{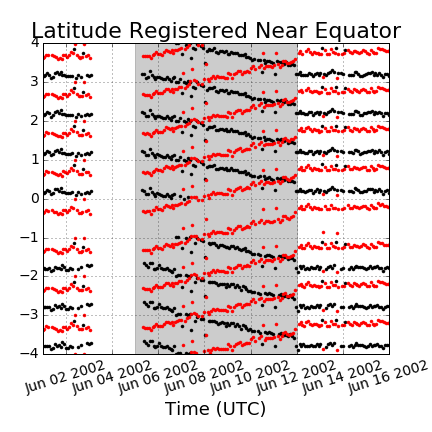}
\caption{(Color online) Latitude of data near the equator, when the spacecraft is moving Northward (black) and Southward (red).  Two regions affected by this cut are indicated by the shaded bands.}
\label{fig:anom2}
\end{figure}

There were a few data reduction cuts that do not fall under the above categories.  This includes something identified in \cite{Maurice2011} as erroneous latitude registration errors.  This can be seen in Fig.~\ref{fig:anom2}, which shows the latitude (for any longitude) registered near the equator.  The black points indicate when the spacecraft is moving Northward (``up") and the red points when it is moving Southward (``down").  Data in this plot have all other cuts already applied, so gaps in time mostly represent SEP event cuts and the effect of removing the Northward equatorial crossing can also be observed.  The two regions of time cut under this error code are both in 2002, and are also shown in Fig.~\ref{fig:anom2} with the cut region shaded gray.

There was also very rarely an issue with the UTC timestamp showing up at the seemingly incorrect time, affecting 63 data points overall.  For example, several data entries with UTC timestamps indicating 2007 showed up at SLCK values near November 25, 2006.  There are two other data points with a similar issue, one in 2009 and one in 2010.  It is likely the data affected by this anomaly are fine and somehow only the UTC timestamp is erroneous.  However, to be safe these data were removed.

Finally, cuts based on the MONS sensor temperature and high voltage power supply (HVPS) were applied.  Note that there appears to be a mismatch in the mapping of engineering data to files starting April 1, 2004.  Before this date, the engineering data are in the appropriately labeled files, which for the HVPS are the hvps\_mntr\_[1,2].dat files.  From April 1, 2004 through the dataset that we considered, the data corresponding to the HVPS1 and HVPS2 were empirically found to be in the files plus\_5v\_crnt\_dig.dat and plus\_5v\_anlg.dat, respectively.  The other files in the eng directory are similarly mis-mapped. We did not determine the mapping for the other engineering data.

\begin{figure}[h!]
\centering
\includegraphics[width=0.48\textwidth]{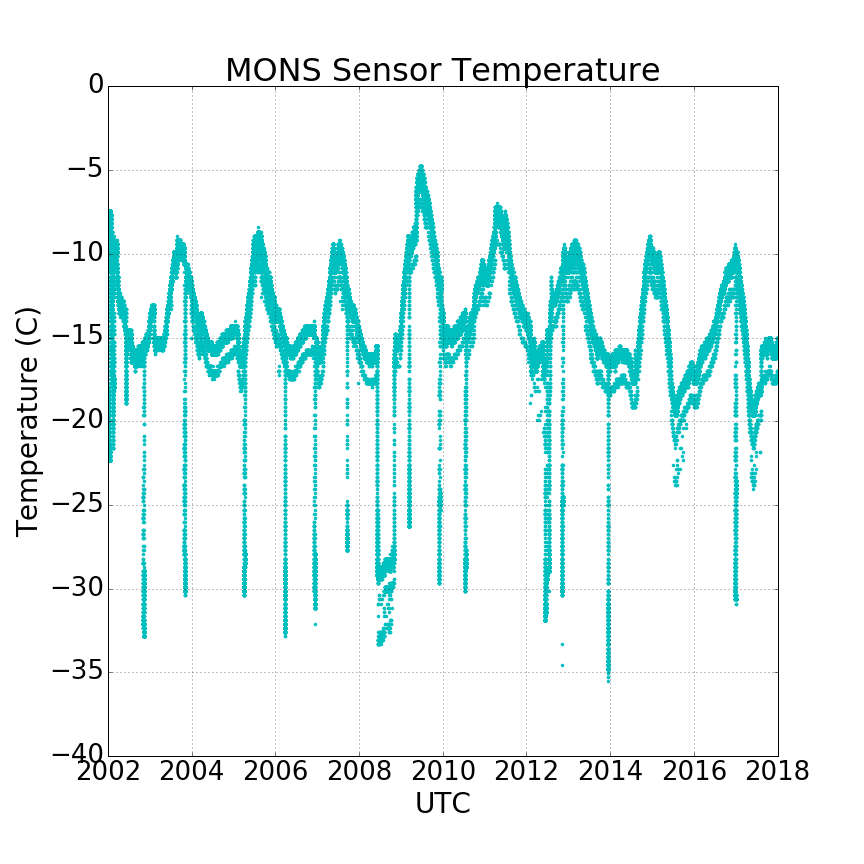}
\includegraphics[width=0.48\textwidth]{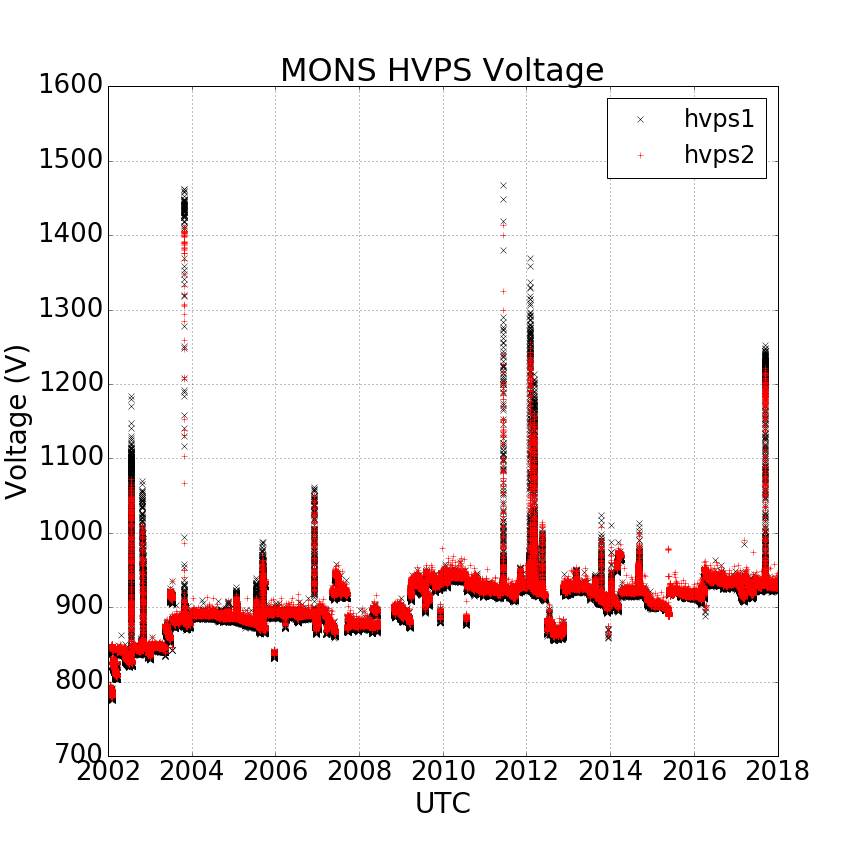}
\caption{(Left) MONS sensor temperature. (Right) HVPS voltage for 2002 -- 2017.}
\label{fig:engdata}
\end{figure}

The temperature and HVPS voltage are shown for the beginning of data collection through the end of 2017 in Fig.~\ref{fig:engdata}.  These data are registered at different SLCK clock values than the rest of the MONS data therefore no event cuts are applied in these plots.  The temperature fluctuates seasonally with some excursions to lower temperature.  The origin of the double-banded structure in the temperature data is not known at this time.  The two HVPS channels track together with some excursions to high voltages.  All of the HVPS excursions occur during SEP events and are therefore excluded based on that cut.  Some of the temperature excursions partially overlap an SEP event, but those that are not are removed.

\subsubsection{Summary of data cuts}

From February 22, 2002 through the end of 2017 there are just over 23 million data points processed.  The total percentage of data removed from all cuts is 14.1\%.  The majority of this, 12.1\% of the data or 85.6\% of the total cut, comes from the removal of SEP events.  Stability cuts remove only an additional 0.2\% of the data.  Spacecraft cuts affect 1.8\% of the data, but 12\% of the data under this cut overlap with SEP event ranges.  Within the spacecraft cuts, 0.56\% of the data is cut due to bad pointing/intersecting flags, 0.4\% of the data exhibits transients in the position or velocity parameters, and 0.28\% of the data are removed from equatorial crossings.  Temperature cuts remove 0.65\% of the data and the erroneous latitude registration early in the dataset affects 0.26\% of the data.

\subsection{Data Corrections}\label{sec:data_correction}

Several data corrections are applied after bad data are removed.  These corrections are necessary to extract the correct prism counting rates and appropriate latitude and longitude registration.

\subsubsection{ADC Non-Linearity}

The analog to digital conversion of the prism spectra introduces differential nonlinearities into the recorded histogram data. To see this best and to determine the correction, Prism 3 histograms were averaged poleward of 85$^{\circ}$N during the Northern summer.  Kilometer-thick perennial water-ice deposits cover most of the region poleward of 80$^{\circ}$N in the Northern summer \cite{Clifford2000}.  A cut selecting the period between solar longitude $L_s = 110^{\circ} - 140^{\circ}$ was used to be safely in Northern summer based on \cite{Piqueux2015b} observations of Northern seasonal cap growth and retreat.  Prism 3 is oriented away from the planet and therefore is expected to show only a smooth, continuous background.  Deviations from this smooth function allow the nonlinearity correction to be determined.

\begin{figure}[h!]
\centering
\includegraphics[width=0.8\textwidth]{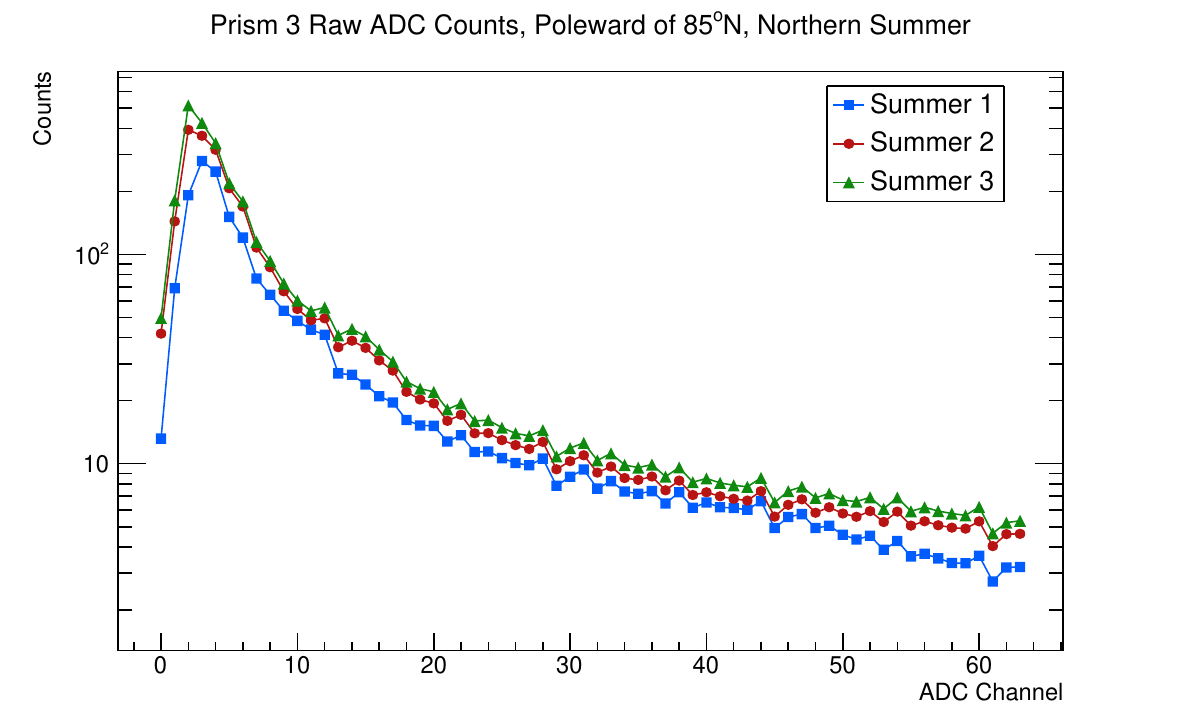}
\caption{Raw ADC counts from Prism 3 averaged over three Northern summer periods.}
\label{fig:adcnonlin1}
\end{figure}

\begin{figure}[h!]
\centering
\includegraphics[width=0.98\textwidth]{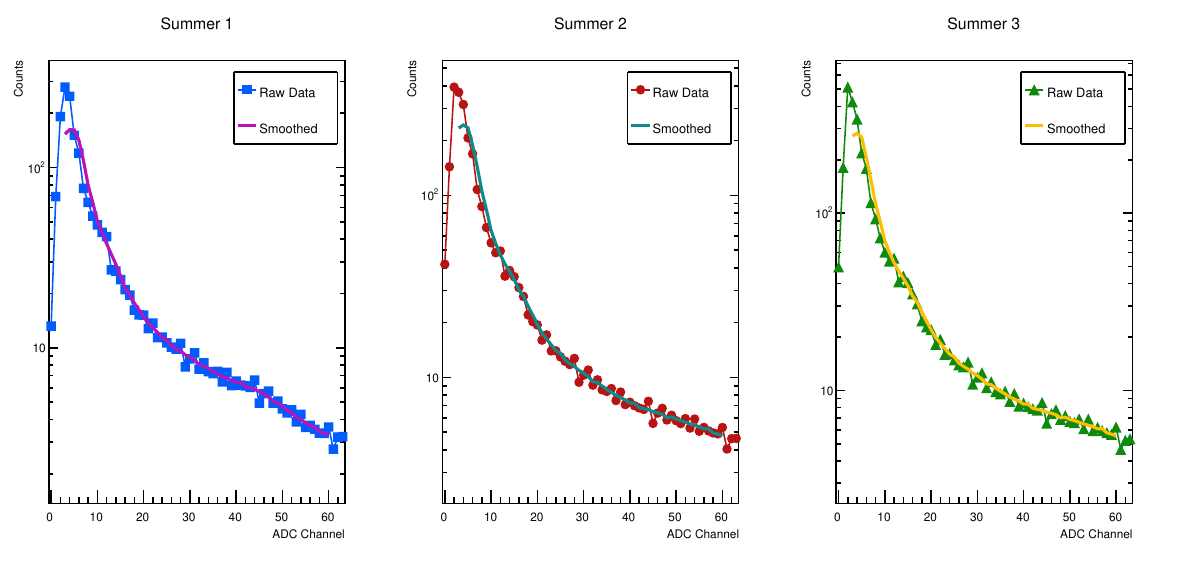}
\caption{Raw ADC counts with smoothed curves from a 7-channel boxcar average.}
\label{fig:adcnonlin2}
\end{figure}

The dataset analyzed for determining the nonlinearity correction (2002--2007) covers three Northern summers within this $L_s$ range that span $\sim$63 days each: 12/18/2002 -- 2/19/2003, 11/4/2004 -- 1/6/2005, and 9/22/2006 -- 11/23/2006.  The raw Prism 3 spectra for these three summers are compared in Fig.~\ref{fig:adcnonlin1}.  
There are some differences year-to-year, likely due to slight differences in sensor temperature and GCR flux.  However, overall a repeating 16-channel pattern can be observed, a result of the ADC design used for the Category 1 histograms.  The data were smoothed by applying a centered 7-channel boxcar filter.  This means the first three channels and last three channels are not included in the smoothed data, which are shown in Fig.~\ref{fig:adcnonlin2}.

\begin{figure}[h!]
\centering
\includegraphics[width=0.98\textwidth]{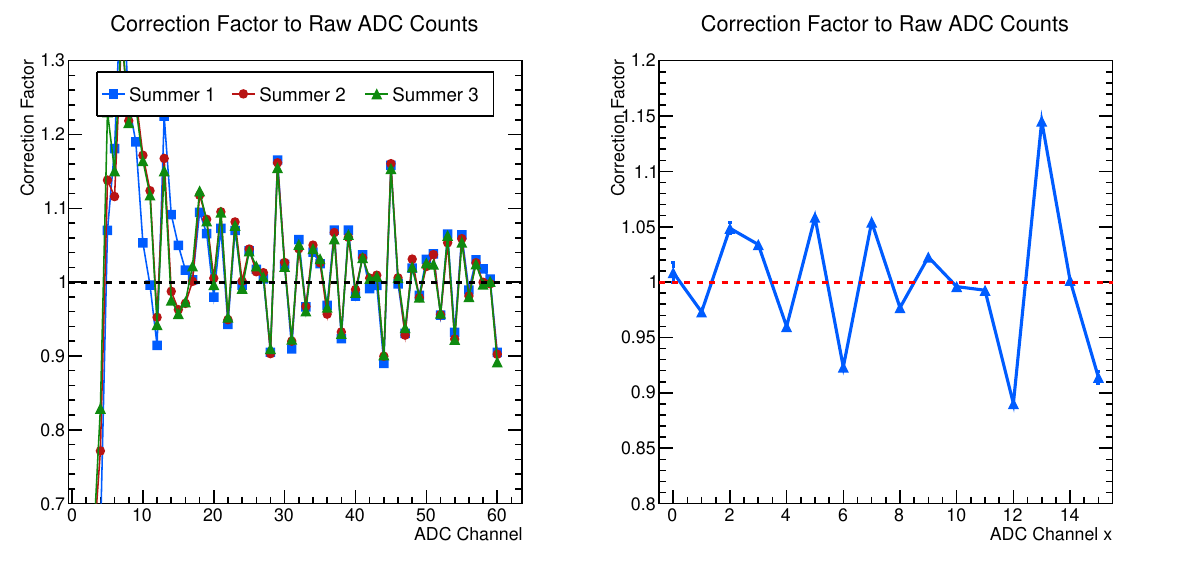}
\caption{Correction factor for ADC nonlinearity. See text for details.}
\label{fig:adcnonlin3}
\end{figure}

The correction factor was determined by calculating 1 + (Smoothed[i] - Raw[i])/Raw[i] for these 58 ADC channels.  The extracted correction factors are shown in the left panel of Fig.~\ref{fig:adcnonlin3}.  To determine the final correction factor, data from the appropriate channels above channel 16 are averaged (\textit{e.g.} channels 16, 32, and 48).  Channels below 16 were excluded to avoid any artifacts from the boxcar smoothing near the peak.  The final correction factor for each set of 16 channels comes from averaging the results of the three summers and is shown in the right of Fig.~\ref{fig:adcnonlin3} and given in Table~\ref{table:adcnonlin}.

\begin{table}[h]
\caption{ADC Nonlinearty correction factor, repeated for each 16 channel group.}
\label{table:adcnonlin}
\centering
\begin{tabular}{|c|c|}
\hline
Channel & Correction Factor \\
\hline
0x	& 1.00864 \\
1x	& 0.97306 \\
2x	& 1.04867 \\
3x	& 1.03395 \\
4x	& 0.95991 \\
5x	& 1.05862 \\
6x	& 0.92326 \\
7x	& 1.05427 \\
8x	& 0.97688 \\
9x	& 1.02267 \\
10x	& 0.99594 \\
11x	& 0.99265 \\
12x	& 0.89037 \\
13x	& 1.14551 \\
14x	& 1.00161 \\
15x	& 0.91398 \\
\hline
\end{tabular}
\end{table}

\begin{figure}[h!]
\centering
\includegraphics[width=0.75\textwidth]{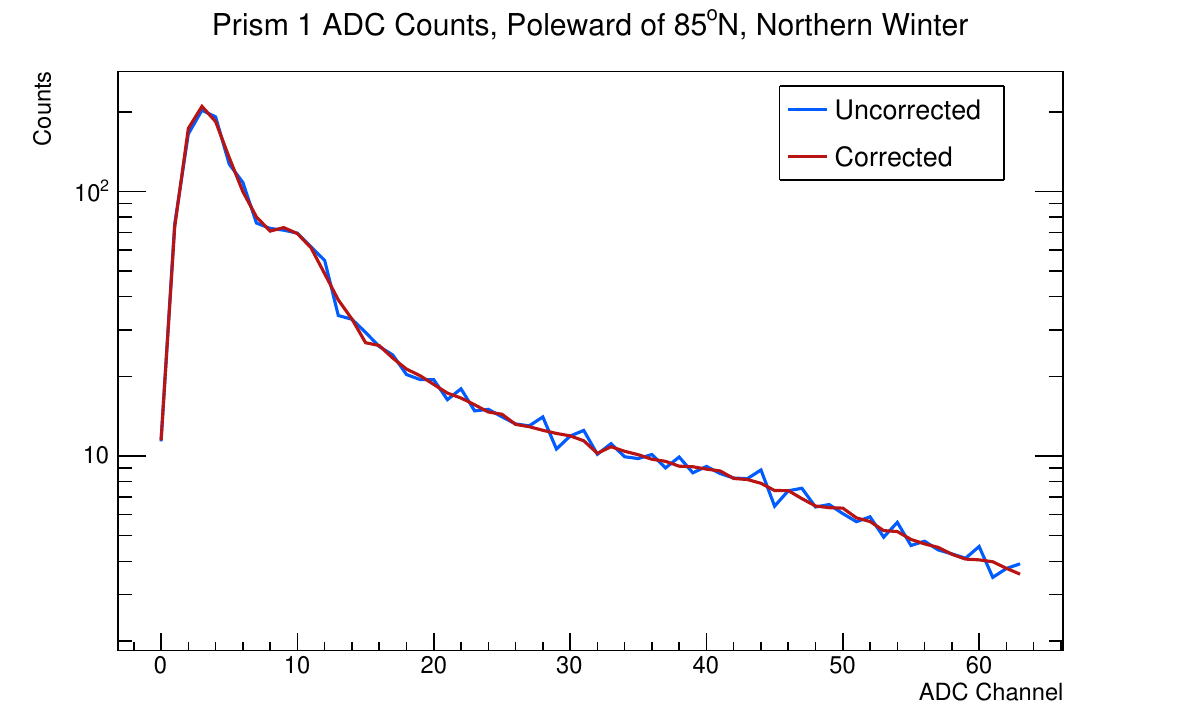}
\caption{Prism 1 ADC counts during a Northern winter, uncorrected and corrected.}
\label{fig:adcnonlin5}
\end{figure}

Figure~\ref{fig:adcnonlin5} shows an example of the ADC nonlinearity correction applied to Prism 1 counts.  The data come from $\sim$12 days during the peak of Northern winter ($L_s = 266^{\circ}-274^{\circ}$) and are averaged over this time period for latitudes poleward of 85$^{\circ}$N.  The total number of counts is increased by 0.7\%, consistent with the average correction factor.  We expect other systematic errors to be much larger than this and therefore are not concerned.

\subsubsection{Gain Correction}

The gain of each prism drifts during the duration of the mission, due to degradation over time, high voltage, or temperature variations.  To determine the gain correction the position of the $^{10}$B neutron capture peak must be identified.  Unless there are abrupt changes in the prism high-voltage, the peak position is very stable and changes smoothly with time.  To acquire good statistics on the determination of the peak location, and the shape of the background that is used later, entries within a 10$^{\circ}$ $L_s$ bin were subdivided into 40 subbins.  The peak location and background parameters were determined on the histograms summed within these subbins, which typically contained approximately 1500 entries.

\begin{figure}
	\centering
	\includegraphics[width=0.49\textwidth]{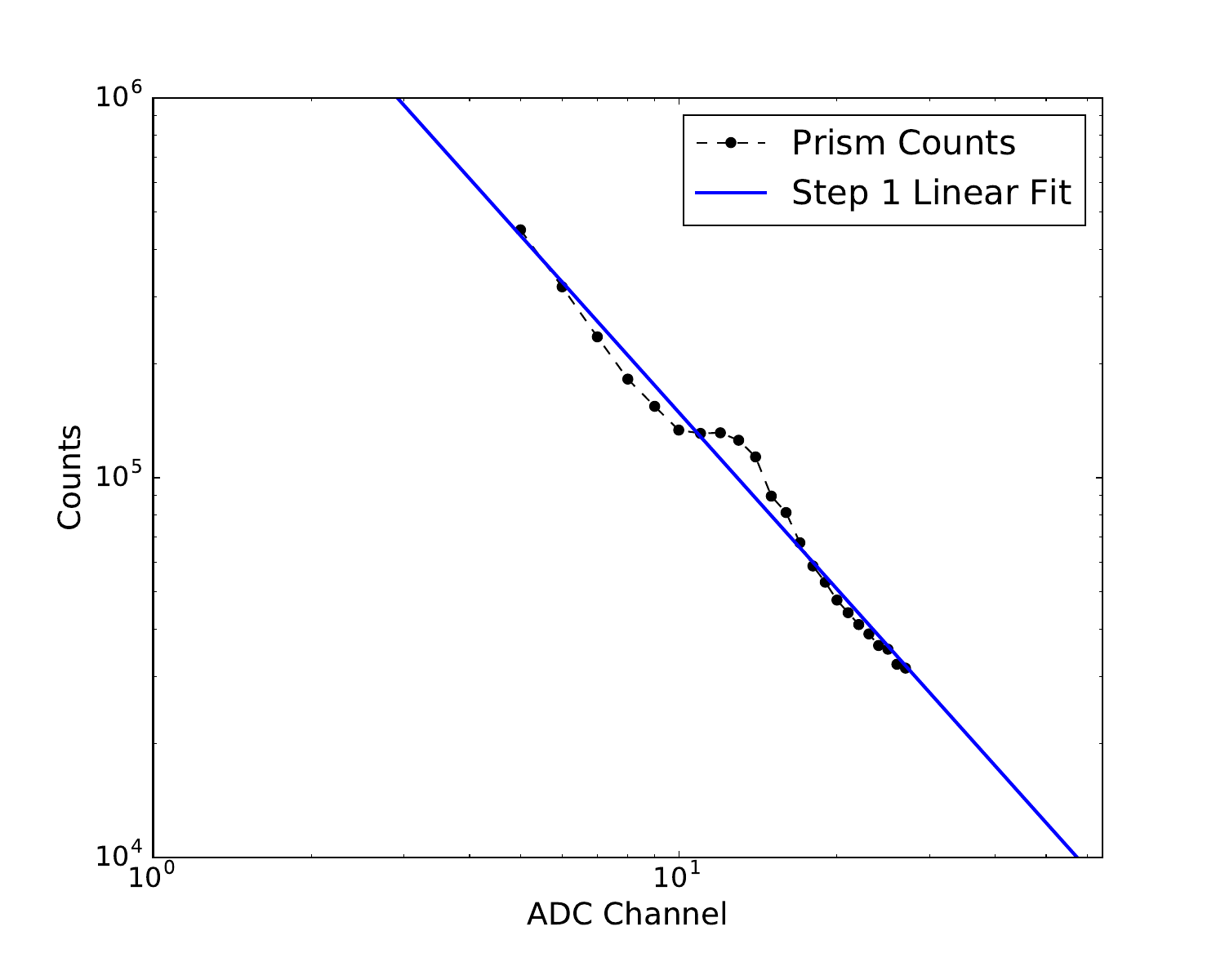}
	\includegraphics[width=0.49\textwidth]{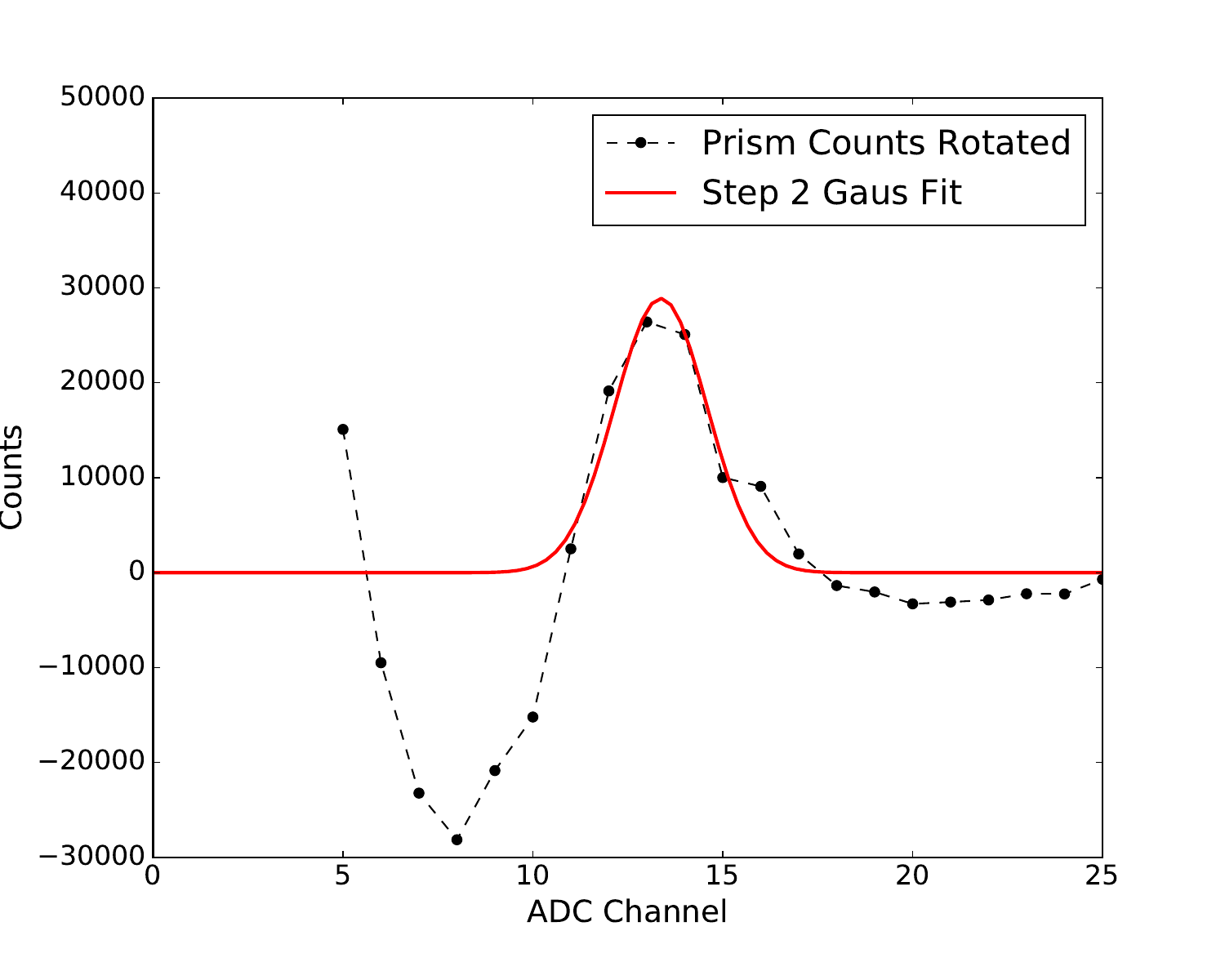}
	\caption{Example illustrating Steps 1 and 2 of the fitting procedure to determine peak position.}
	\label{fig:gain2}
\end{figure}
The fitting procedure used in this analysis includes 5 steps and is illustrated using Prism 4.  Steps 1 and 2 (Fig.~\ref{fig:gain2}) follow similarly to what was done in \cite{Maurice2011}; a linear fit in log-log space is performed and then a Gaussian is fit to the rotated data.  In these steps only part of the spectrum is fit, from a minimum ADC channel as determined by the maximum value plus an offset (this safely places the minimum fit after the low-channel roll over) to a maximum ADC channel of channel 28.  The offset was prism- and time-dependent, as the gain of the neutron capture peak varies, with typical values between 1--3 channels.  The maximum ADC channel of 28 was chosen to be high enough above the peak to ensure a good fit but below a background feature observed at high ADC channels.  The initial guess for the mean value of the Gaussian fit in Step 2 is determined by finding all the local maxima in the rotated array and requiring that the mean value is not immediately within a certain number of channels to the minimum fit location and of the remaining candidates is selected as the one with the highest counts value.  This results in a single and most often correct guess for the mean value.  In Step 3 the Gaussian fit mean and sigma are used to create an excluded region around the peak of $\pm$3$\sigma$ from the mean, and the linear fit is repeated over the same constrained minimum and maximum ADC fit values in an attempt to improve the linear fit parameters.  Step 4 repeats the rotation based on the tuned linear fit and refits a Gaussian to extract tuned Gaussian parameters.  The tuned Gaussian mean is the peak location and is used to apply the gain correction that lines up all data to have a peak in channel 10.

\begin{figure}[h!]
\centering
\includegraphics[width=0.98\textwidth]{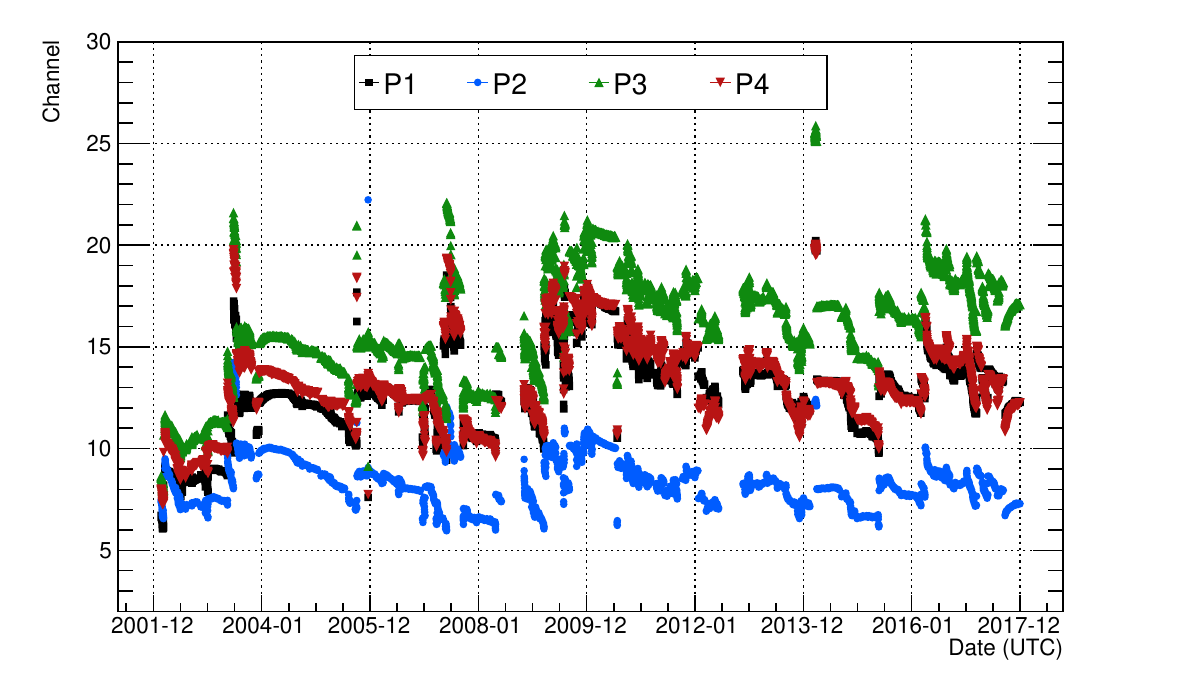}
\caption{Channel location of the $^{10}$B capture peak for each prism over all processed data from 2002 -- 2017.}
\label{fig:gain4}
\end{figure}

The peak positions for the four prisms from the start of the mission through the end of 2017 are shown in Fig.~\ref{fig:gain4}.  There is a period of approximately 3.5 months during MY 31 (8/4/2012 -- 11/16/2012) that the prism gains are too low and peak values cannot be reliably extracted; this data has been removed.

\subsubsection{Peak Integration}

\begin{figure}[b!]
\centering
\includegraphics[width=0.65\textwidth]{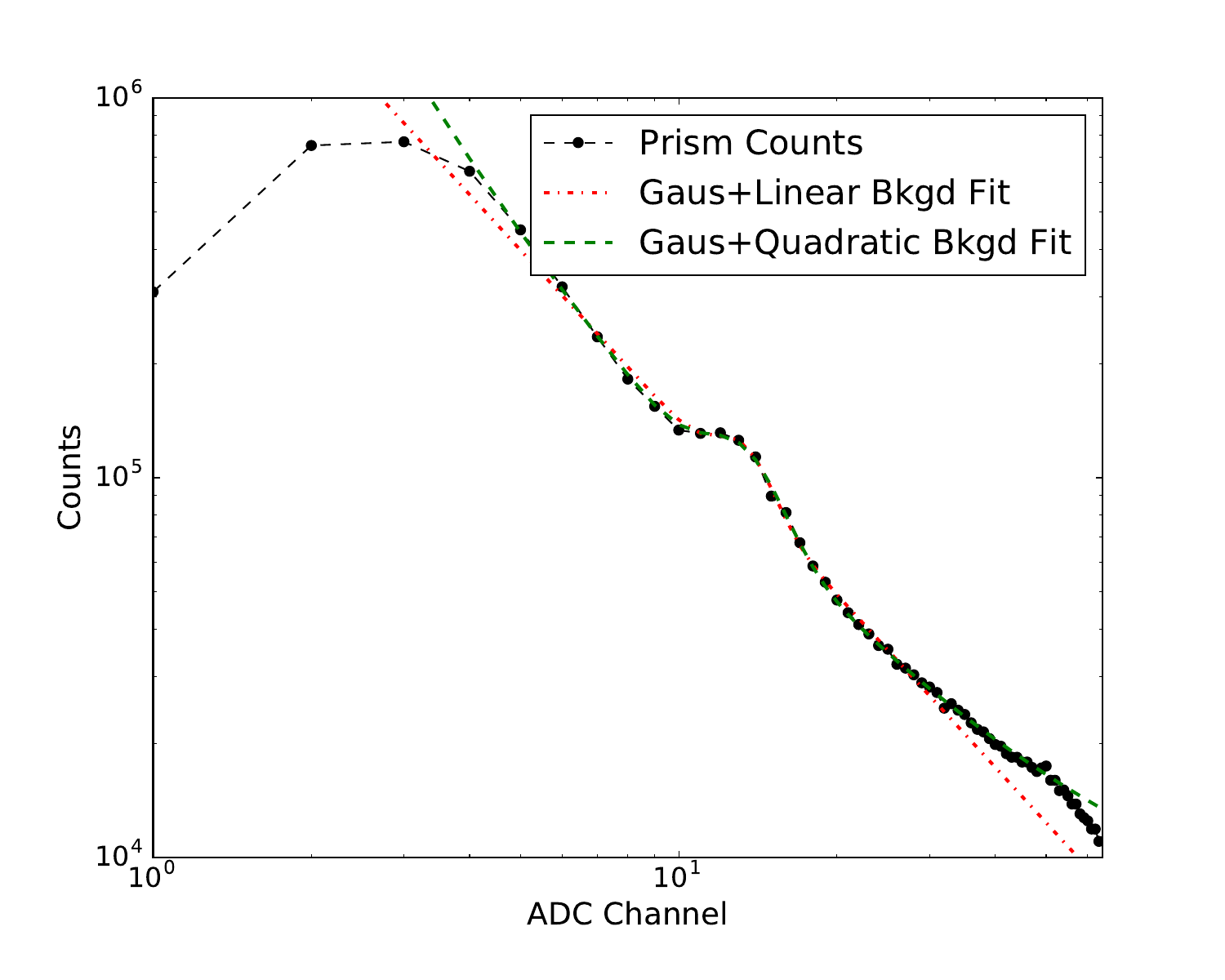}
\caption{Example illustrating background fits.}
\label{fig:gain3}
\end{figure}
Once the histograms have been gain corrected, background parameters can be fit and the histograms integrated to determine the signal counts.  The background fit parameters were constrained by using the same summed histograms as used to determine the gain correction.  The tuned log-log linear fit parameters and the gain-corrected Gaussian fit parameters from Steps 3 and 4 above are used to guide initial guesses of a Gaussian plus background fit of the spectrum in log-log space.  The results are compared for a linear background and a quadratic background in Fig.~\ref{fig:gain3}. Both background fits yield essentially the same mean peak position, however, the quadratic background fits the curve better and yields a better background subtraction, in particular when the peak location is high within the ADC range.  When fitting each individual data point, the Gaussian mean, Gaussian sigma, background slope, and 2$^{nd}$-order background polynomial coefficient are constrained by knowledge of the higher-statistics fits, while the Gaussian height and background offset are allowed to float.  For each data point, this new fit results in parameterizations for the signal and background functions, which are used to integrate and determine the number of signal counts.  No deadtime correction is made, as \cite{Maurice2011} showed this is quite small, and therefore all counts are simply divided by 19.75~s to determine the count rate per second.

\subsubsection{Altitude Correction}

\begin{figure}[h!]
\centering
\includegraphics[width=0.6\textwidth]{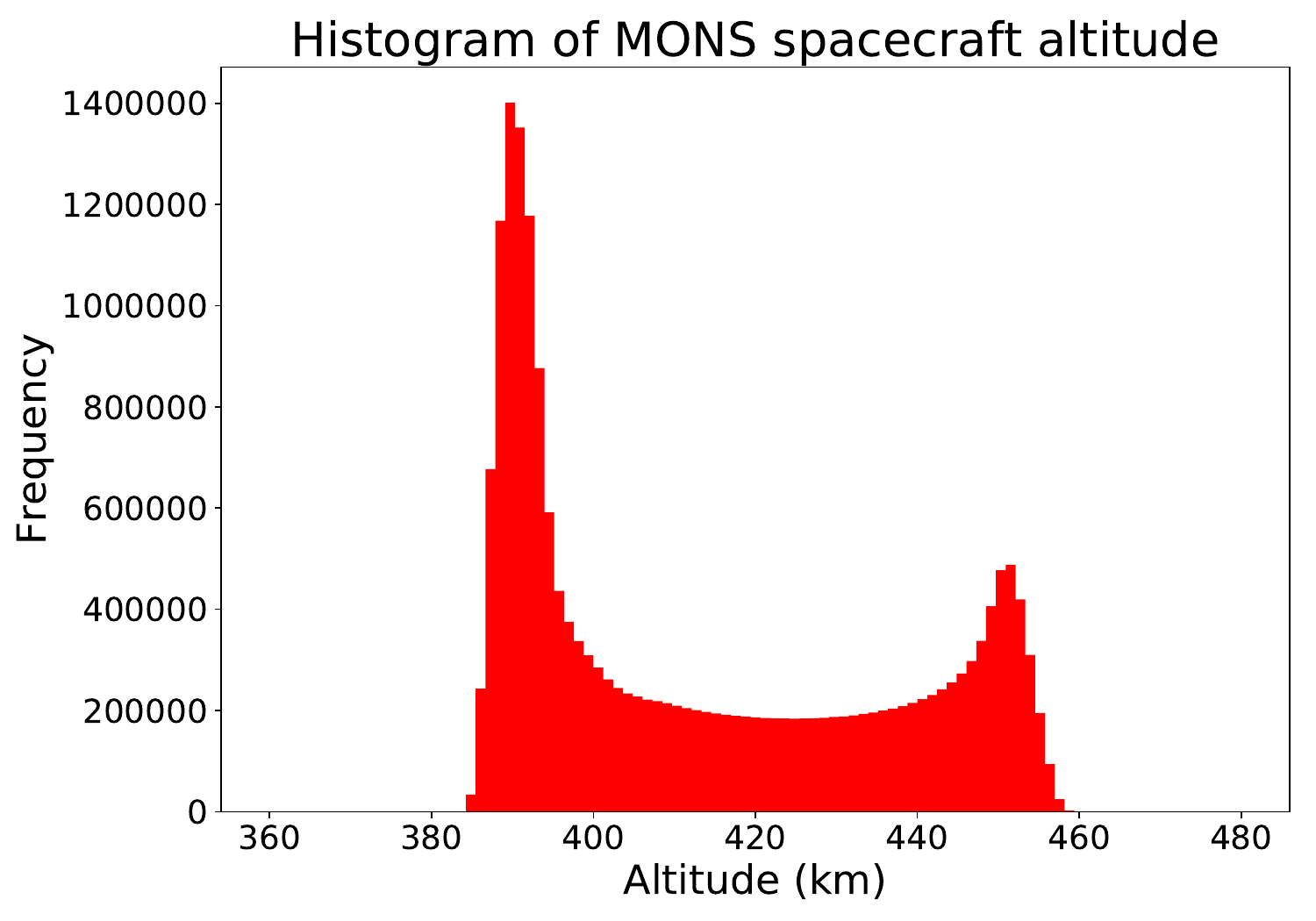}
\caption{Altitude of MONS with respect to the mean Mars surface sphere.}
\label{fig:altitude}
\end{figure}

The orbit of MONS is slightly elliptical, with an altitude that changes from $\sim$380--460~km.  Figure~\ref{fig:altitude} shows a histogram of the spacecraft altitude for all of the processed data from the start of the mission in 2002 through 2017.  The counting rates are normalized to an altitude of 400~km using the following equation which corrects for the solid angle observed at a given altitude $h$ \cite{PrettymanBook}:
\begin{equation}
\Omega(h) = 2\pi \left[1 - \sqrt{1 - R^2/\left(R+h\right)^2}\right]~.
\end{equation}
where $R = 3389.5$~km is the mean radius of Mars.  The scale factor is calculated as $\Omega$(400~km)/$\Omega$(h) and varies from $\sim$0.99--1.05.  We do not make any corrections for local elevation of the surface.

\subsubsection{Ground-Track Correction}

\begin{figure}[h!]
\centering
\includegraphics[width=0.49\textwidth]{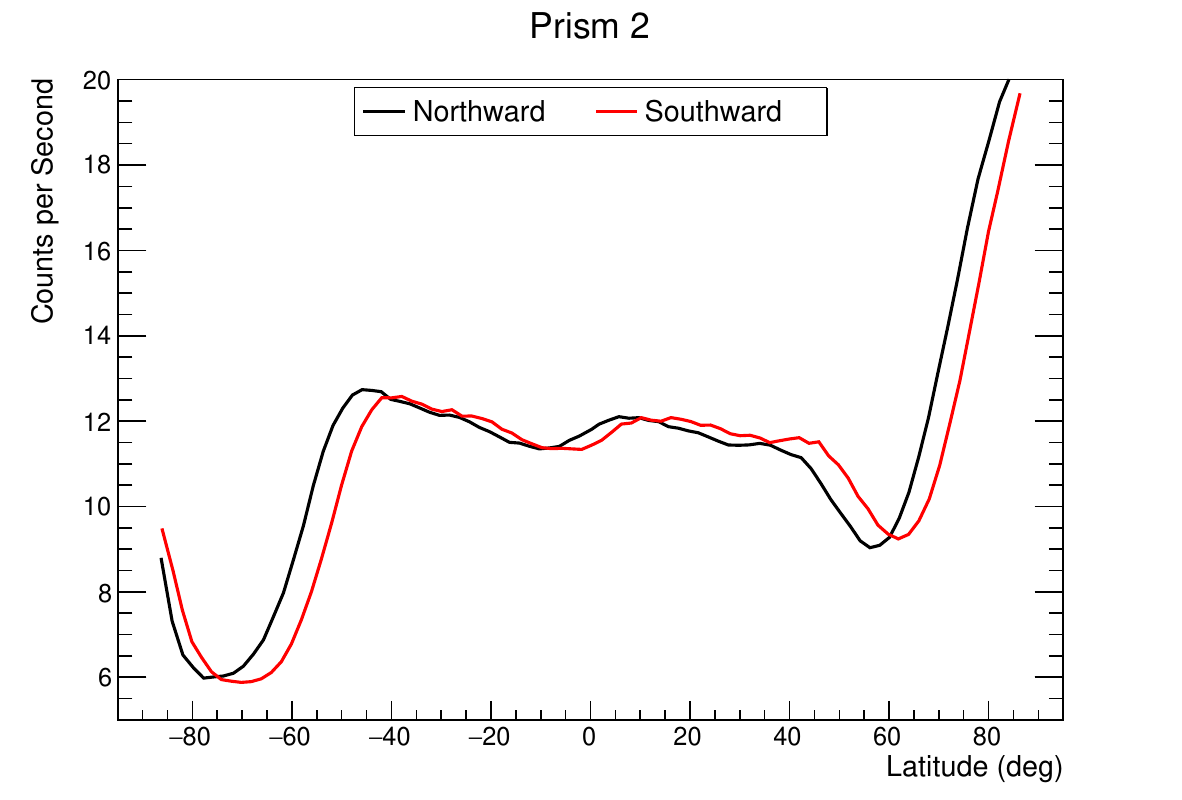}
\includegraphics[width=0.49\textwidth]{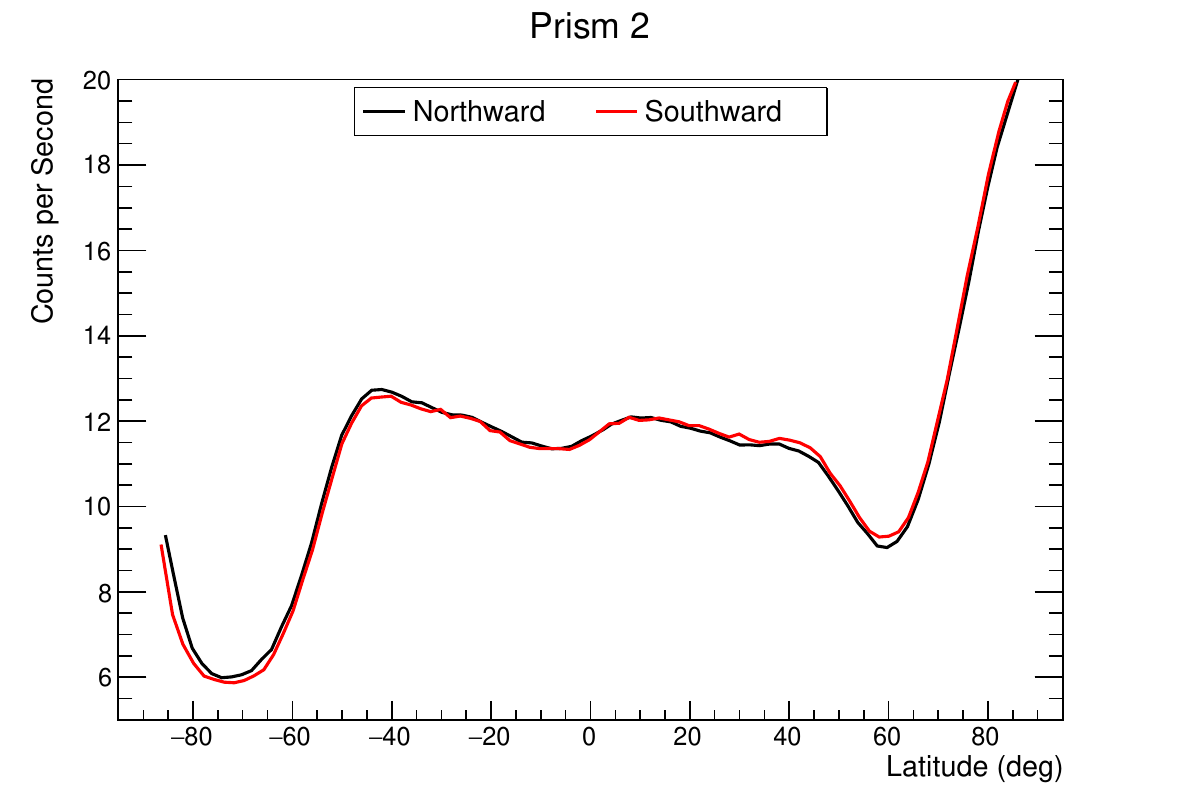}
\caption{Prism 2 Northward and Southward counts averaged over all longitudes for the first quarter of 2004, showing an angle offset originating from spacecraft motion (left panel) and its correction (right panel).}
\label{fig:latoffset_1}
\end{figure}

The MONS detector is positioned such that Prism 1 faces nadir and Prisms 2 and 4 are along and opposite to the direction of spacecraft motion, respectively.  Because of the motion of the spacecraft and the fact that thermal and epithermal neutrons have similar velocity to the spacecraft, the latitude and longitude that each prism ``sees'' (\textit{i.e.} ground track) is not the same as the sub-spacecraft location that is registered with each data point.  To determine the angle offset between the registered location and the observed location, the counting rates summed over longitude as a function of latitude when the spacecraft is traveling Northwards and Southwards can be compared.  An observable shift is seen in the Northward versus Southward data, as illustrated in the left panel of Fig.~\ref{fig:latoffset_1} using data from the first quarter of 2004.  The angle that resolves this discrepancy is the angle offset, and this depends on the prism.  As the spacecraft orbit has an inclination of 93.2$^{\circ}$, both the latitude and the longitude have to be corrected.

\begin{figure}[h!]
\centering
\includegraphics[width=0.65\textwidth]{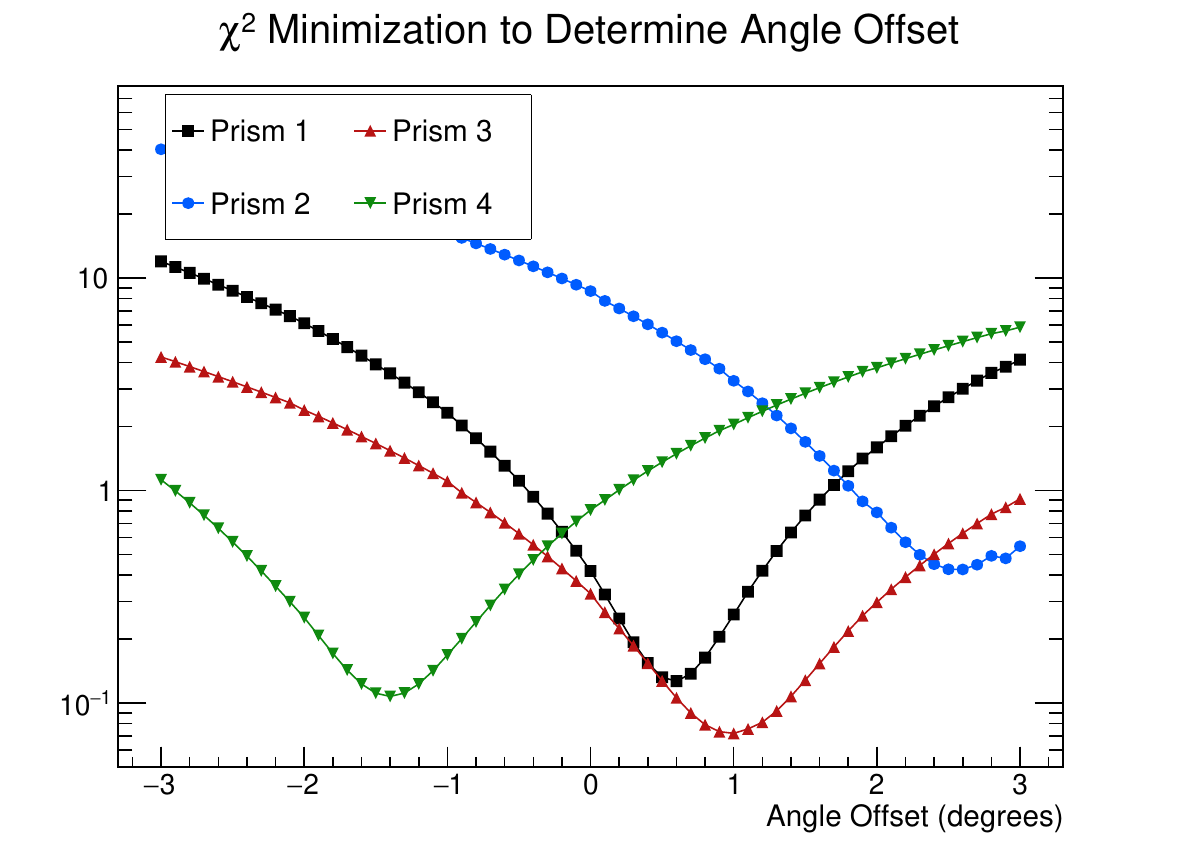}
\caption{$\chi^2$ optimization to determine prism angle offsets.}
\label{fig:latoffset_2}
\end{figure}

To determine the angle offset that best resolves the discrepancy for each prism, a $\chi^2$-minimization was performed which is shown in Fig.~\ref{fig:latoffset_2}.  The resulting angles are 0.4289$^{\circ}$ for Prism 1, 2.5297$^{\circ}$ for Prism 2, 0.8757$^{\circ}$ for Prism 3, and -1.5667$^{\circ}$ for Prism 4.  A positive number indicates the prism is seeing ahead of the spacecraft motion, while a negative number is behind the spacecraft.  These numbers are within 10-20\% to those published in \cite{Maurice2011}.

This correction is applied to each data point where we first calculate the bearing (direction) of the ground track by using the latitude and longitude of the current and subsequent data points.  Given the latitudes $\theta_{1,2}$ and longitudes $\varphi_{1,2}$ of the initial point (1) and final point (2), the bearing $\lambda$ is given by:
\begin{equation}
\lambda = \textrm{atan2}\left(\sin\left(\varphi_2-\varphi_1\right)\cos\theta_2,\cos\theta_1\sin\theta_2 - \sin\theta_1\cos\theta_2\cos\left(\varphi_2-\varphi_1\right)\right)~.
\end{equation}
This is combined with the knowledge of the angle offset determined from the $\chi^2$ optimization ($\delta_i$ for Prism $i$) to calculate the new latitude and longitude to register with each prism:
\begin{eqnarray}
\theta^i_{\textrm{new}} &=& \textrm{asin}\left(\sin\theta_1\cos\delta_i + \cos\theta_1\sin\delta_i\cos\lambda\right)~, \\
\varphi^i_{\textrm{new}} &=& \varphi_1 + \textrm{atan2}\left(\sin\lambda\sin\delta_i\cos\theta_1,\cos\delta_i-\sin\theta_1\sin\theta^i_{\textrm{new}}\right)~.
\end{eqnarray}
Applying these corrections to the first quarter of 2004 shows a resolved offset in the Northward and Southward counting rates versus latitude, as shown in the right panel of Fig.~\ref{fig:latoffset_1}.

Rotations of the spacecraft about the velocity direction were performed during several periods in 2009 -- 2011. This does not impact Prisms 2 and 4 which are oriented along the direction of motion, however, data from Prism 1 has been removed during these periods since it does not face nadir.   The rotations occur between MY 29, $L_s\approx265$ through MY 30, $L_s\approx34$ and between MY 30 $L_s\approx316$ through MY 31, $L_s\approx61$.

\subsubsection{GCR Correction}

The GCR flux, which is the source for the measured neutron signals, varies over time mostly with solar cycle.  To remove this effect, the data must be normalized to a particular date with a known GCR flux.  To determine the GCR correction factor, the belly band procedure \cite{Maurice2011} is adopted, which assumes that near the equator the ground-surface processes are in equilibrium and therefore the neutron counting rates should be stable over time.  In addition to the GCR flux changing over time, seasonal changes in the density of Mars' atmosphere can lead to seasonal changes in the neutron counting rates.  These must be accounted for through simulations before the GCR correction can be determined.  A radiation transport tool to simulate the neutron leakage flux from Mars and a tool to transport this flux to the MONS spacecraft and predict the prism counting rates were developed.  The details of these tools are discussed in the Appendix.  

We divided the MONS data into $2^{\circ} \times 2^{\circ}$ latitude and longitude bins within $\pm$20$^{\circ}$ of the equator.  The Mars Climate Database (MCD) v5.3 global circulation model (GCM) \cite{Forget1999} was used to determine the atmospheric density on the latitude and longitude grid as a function of seasonal $L_s$.  The neutron counting rates were normalized to an average atmospheric density of 16~g/cm$^2$, using the simulation described in the Appendix.   With seasonal effects from the atmosphere removed from the data, the GCR correction was then determined.

We chose to normalize the GCR proxy to be unity in June 2008.  This corresponds to a solar modulation of $\phi = 463$~MV according to the latest Usoskin model \cite{Usoskin2017} and was chosen because it is during the time period with the lowest uncertainty in the determination of the solar modulation.  This is different than \cite{Maurice2011} where the data were normalized to the period October -- November 2002.  Based on the difference in solar modulation and therefore integrated GCR flux, we expect the counting rates presented in this work to be up to a factor of 2 higher than the counting rates determined in \cite{Maurice2011}.

\begin{figure}[h!]
\centering
\includegraphics[width=0.98\textwidth]{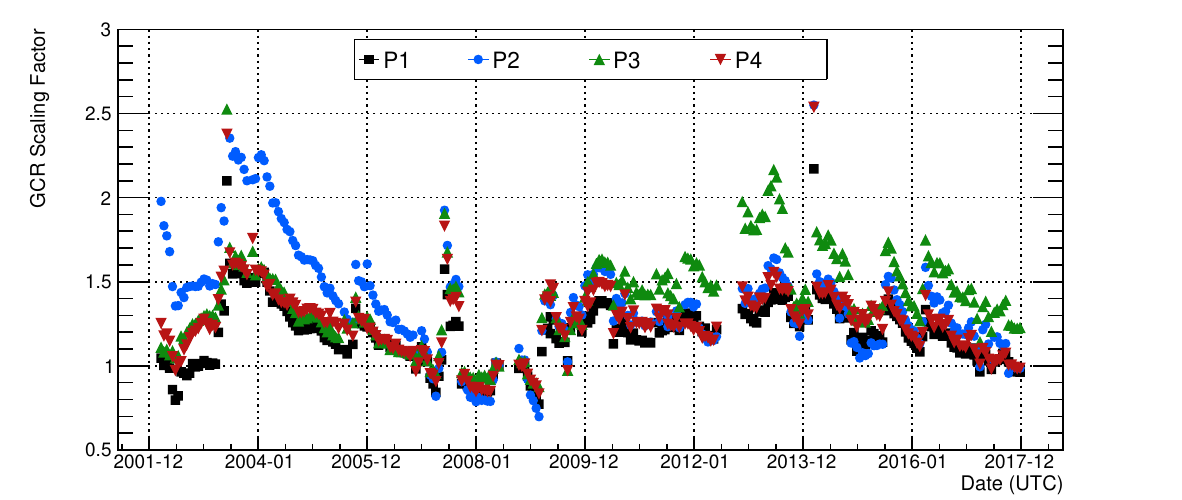}
\caption{GCR normalization relative to June 2008.}
\label{fig:gcrnorm}
\end{figure}

The multiplicative GCR correction is shown in Fig.~\ref{fig:gcrnorm}.  The GCR correction is prism- and time-dependent, as it depends on each prisms efficiency and response to the time-varying GCR flux, which depends on the prism gain.  Since the chosen GCR normalization date is close to solar minimum, when the GCR flux is largest, the GCR correction factor is generally greater than 1.  Between solar minimum and solar maximum, the GCR flux can change by a factor of 2.5.

\section{Results}\label{sec:results}

\begin{figure}[h!]
\centering
\includegraphics[width=1\textwidth]{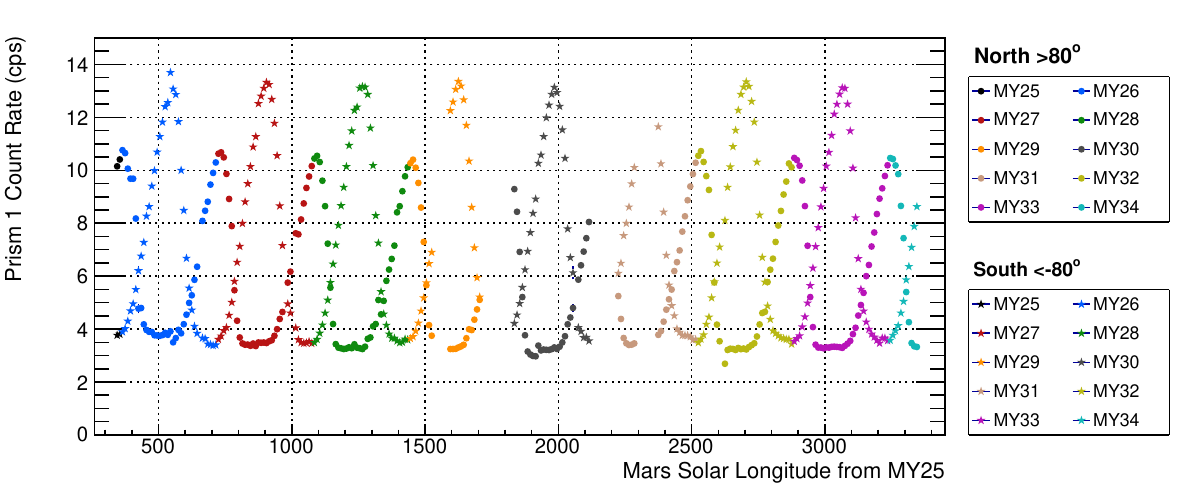}
\includegraphics[width=1\textwidth]{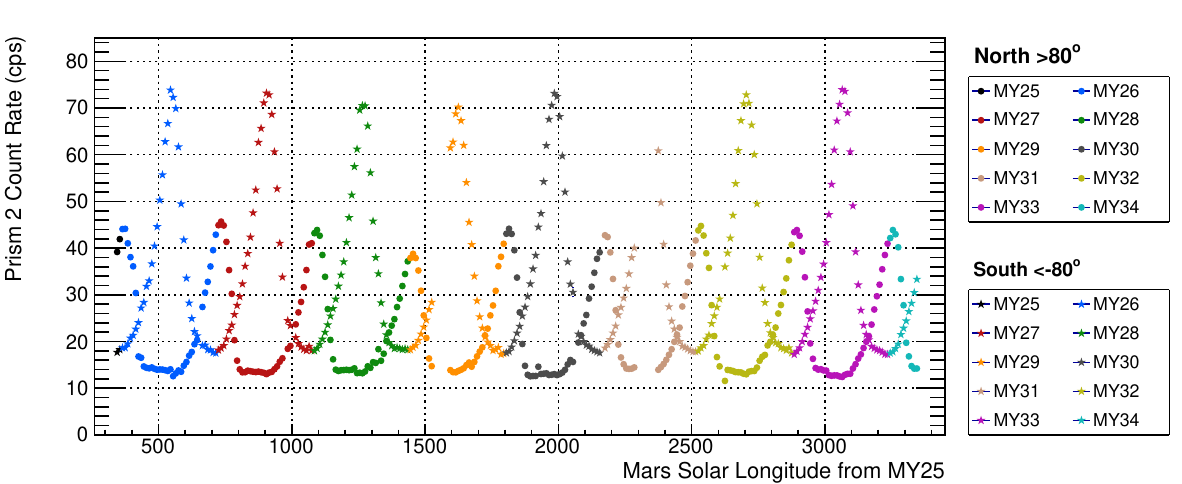}
\includegraphics[width=1\textwidth]{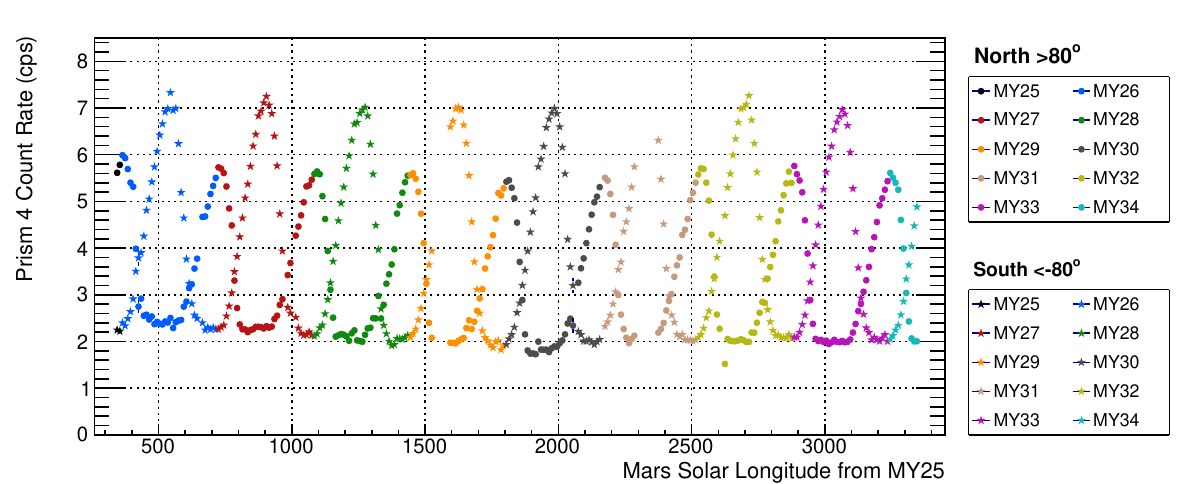}
\caption{Counting rates in 10$^{\circ}$ $L_s$ bins (starting from $L_s$ = 0 in MY 25) for Prism 1 (top) Prism 2 (middle) and Prism 4 (bottom) from the new dataset.  The colors separate the Mars Year, and the circles correspond to data $>$80$^{\circ}$N and the stars data $<-$80$^{\circ}$N.}
\label{fig:result_fig14compare}
\end{figure}

As an example of the temporal coverage available in the new dataset, the prism counting rates in the polar regions are shown in Fig.~\ref{fig:result_fig14compare} for the entire dataset in 10$^{\circ}$ $L_s$ bins.  Data from the North pole (latitude $>$80$^{\circ}$N) is overlayed with the data from the South pole (latitude $<-$80$^{\circ}$N).  These plots can be compared with the results from \cite{Maurice2011} (Figs.~14 and 23), which show similar plots extending part way through MY 29.  As CO$_2$ frost is deposited seasonally the neutron counting rates increase, due to CO$_2$ having a low cross section for absorbing the GCR-induced neutrons, until the seasonal CO$_2$-ice cap reaches peak mass. The counting rates then decreases as the CO$_2$ frost is sublimed away.  

Qualitatively the same trends are observed in this dataset when compared with \cite{Maurice2011}. The baseline count rate in the Northern summer is lower than the Southern baseline count rate, which relates to differences in the regolith and perennial ice caps at each pole.  The counts at the peak of seasonal CO$_2$-frost deposition in the South are higher than the peak counts in the North.  There is also a slight reduction in the Northern peak counts in MY 28--29 relative to previous years observed in Prism 2 that is not seen in Prism 1, similar to observations in \cite{Maurice2011}.  Quantitatively as mentioned previously, differences in the GCR normalization result in the present counting rates being larger than those in \cite{Maurice2011}; on average the Prism 1, 2, and 4 counting rates are a factor of 1.4, 2.6, and 1.7 higher than \cite{Maurice2011}.  The larger difference in the Prism 2 counting rates is attributed to using a quadratic background fit instead of a linear fit in the determination of the counts, as Prism 2 was observed to have the largest 2nd-order polynomial coefficient.  The ratios of peak counting rates in the South to North are generally consistent with \cite{Maurice2011}.

\begin{figure}
\centering
\includegraphics[width=1\textwidth]{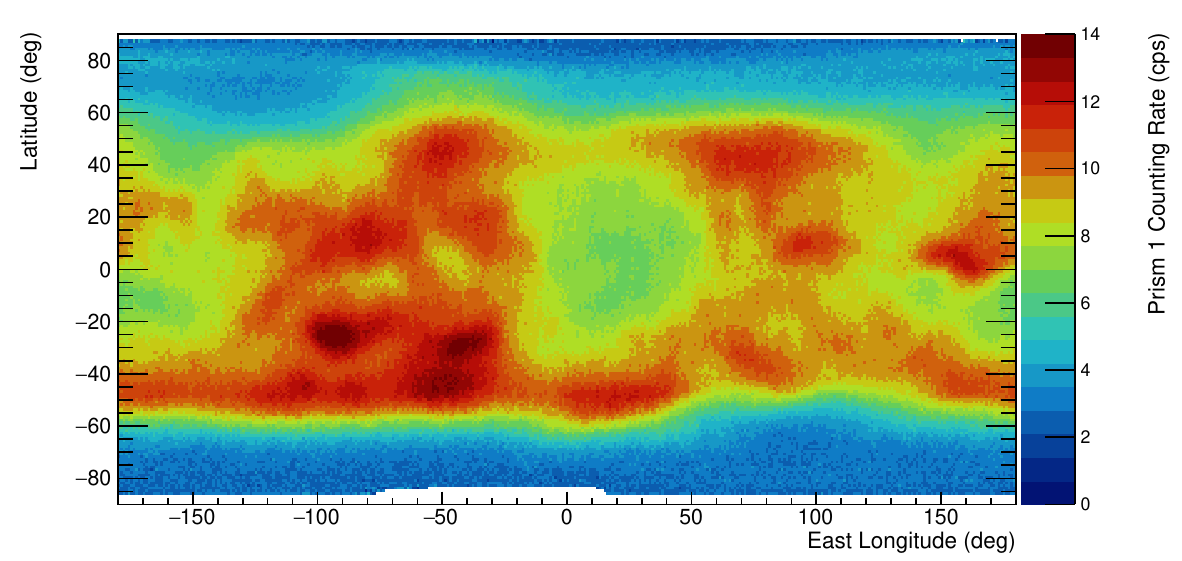}
\includegraphics[width=1\textwidth]{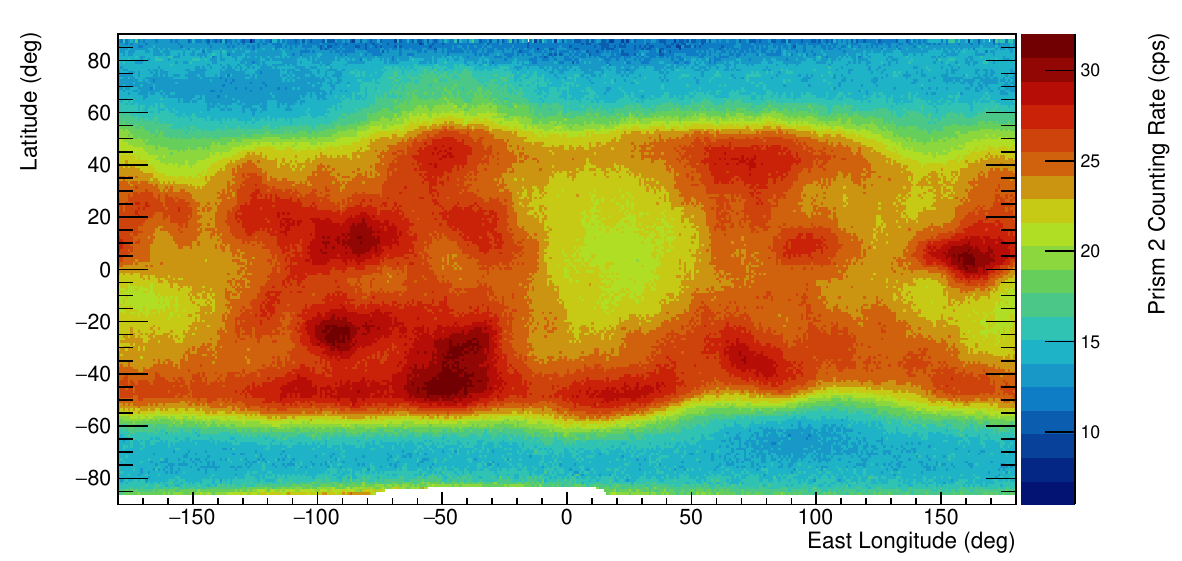}
\includegraphics[width=1\textwidth]{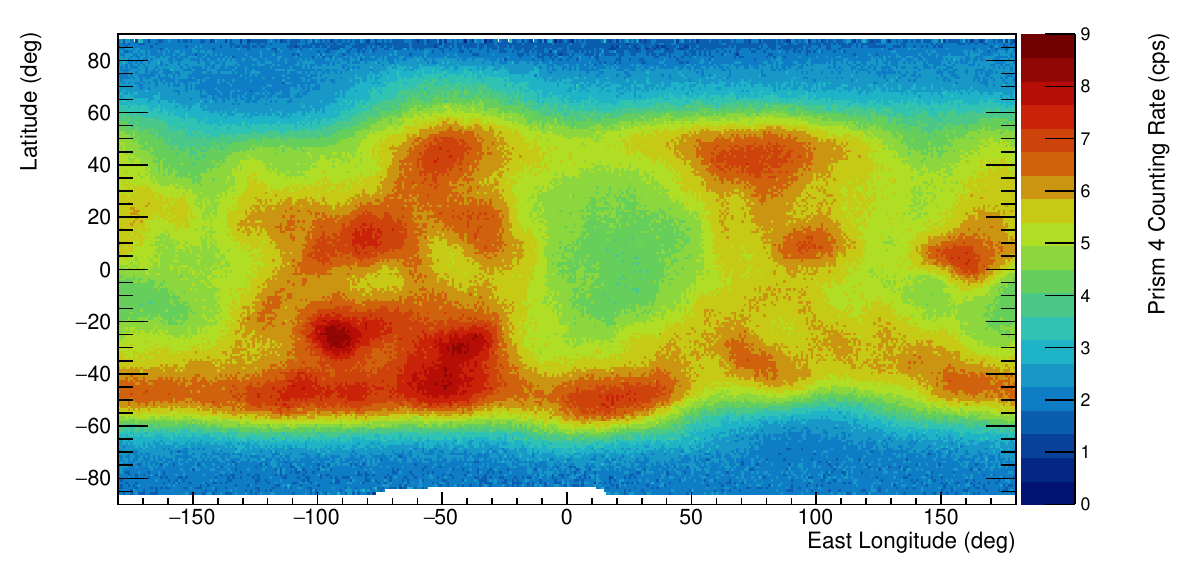}
\caption{Frost-free averaged neutron counting rate maps for Prism 1 (top) Prism 2 (middle) and Prism 4 (bottom) using 1$^{\circ}$ latitude and longitude bins.}
\label{fig:result_fig12compare}
\end{figure}

The new dataset can be averaged over all years to produce averaged neutron counting rate maps for each of the prisms.  These averaged maps are shown in Fig.~\ref{fig:result_fig12compare} assuming 1$^{\circ}$ binning in latitude and longitude.  The data plotted are only CO$_2$ free data, assuming a cutoff of 0.2~g/cm$^2$, similar to \cite{Maurice2011}.  Perennial and seasonal CO$_2$ are not distinguished in this cut, therefore data from the permanent CO$_2$ cap in the South are not included in these plots.  The CO$_2$ frost thickness was predicted for each latitude and longitude bin as a function of $L_s$ using the MCD GCM model \cite{Forget1999}.  These maps are very similar to those found in Fig.~12 in \cite{Maurice2011}. The counting rates in these frost-free maps are inversely proportional to water content in the near-surface.

\begin{figure}[h!]
\centering
\includegraphics[width=0.5\textwidth]{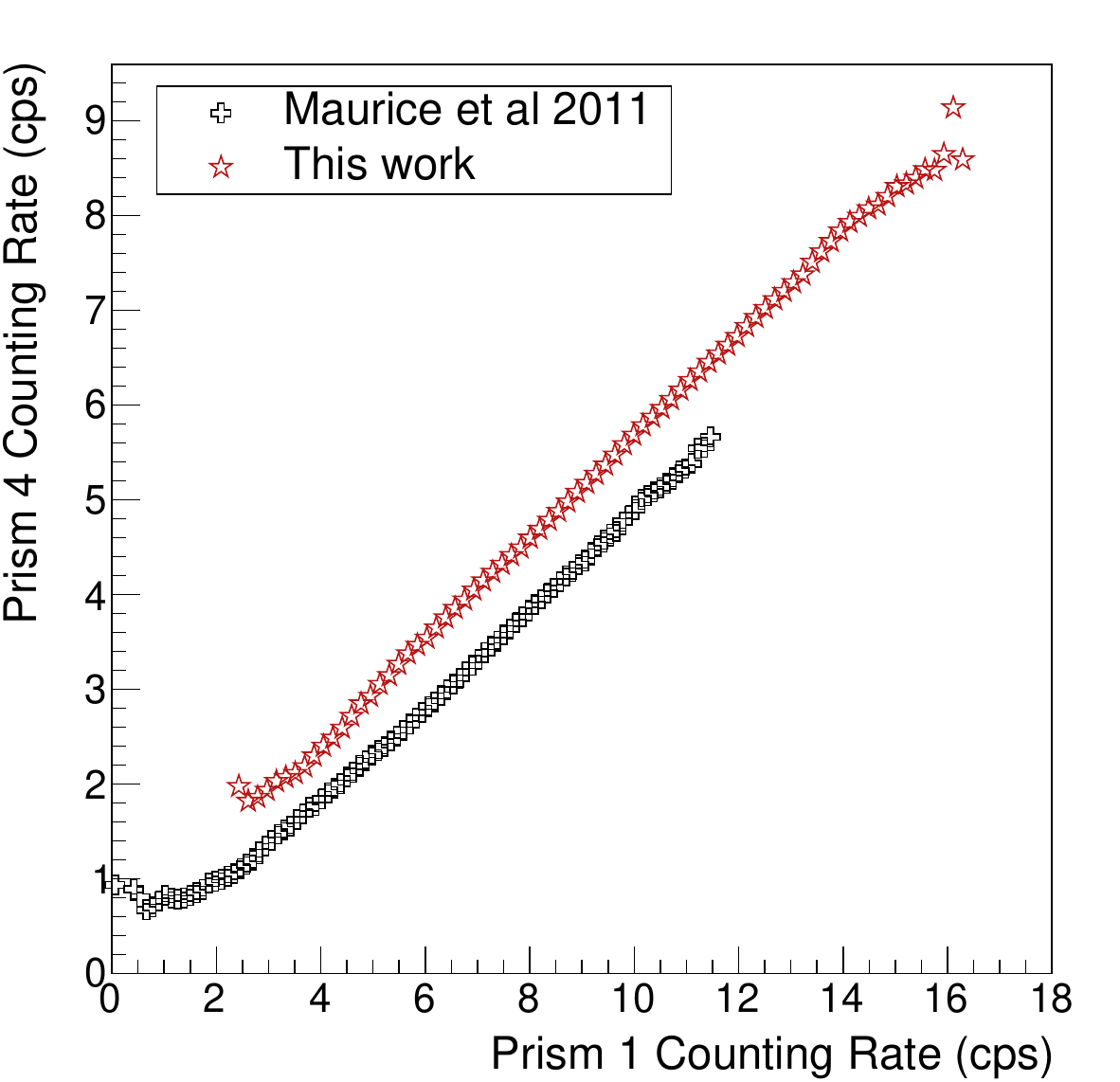}
\caption{Correlation of Prism 4 counting rates to Prism 1 counting rates for frost free data.}
\label{fig:result_p1p4corr_ff}
\end{figure}

Since Prism 1 and Prism 4 provide different measures of epithermal neutrons (and with Prism 1 having a small contamination from thermal neutrons), we plot the correlation of the frost-free counting rates in Prism 4 to the counting rates in Prism 1 in Fig.~\ref{fig:result_p1p4corr_ff}.  There is a strong correlation that is well-fit by a straight line, with a slope of 0.54 in this dataset and a slope of 0.48 from the Maurice \textit{et al.} dataset \cite{Maurice2011}, which is shown for comparison. The effect of differences in the normalization of the data in this analysis is evident.  Since Prism 1 has a larger dynamic range and smaller errors (see discussion of errors and Fig.~\ref{fig:result_fig19compare} below), it is a better choice as the epithermal detector when frost free data are considered.  However, as discussed in \cite{Prettyman2009}, when studying the polar regions the thermal neutron counting rate is extremely sensitive to changes in the atmospheric abundance of N$_2$ and Ar, which can vary seasonally.  Therefore, differences in the epithermal neutron behavior between Prism 1 (which has some thermal neutron sensitivity) and Prism 4 may provide some information on how the atmosphere changes seasonally, and this should be studied in future work.

\begin{figure}[h]
\centering
\includegraphics[width=1\textwidth]{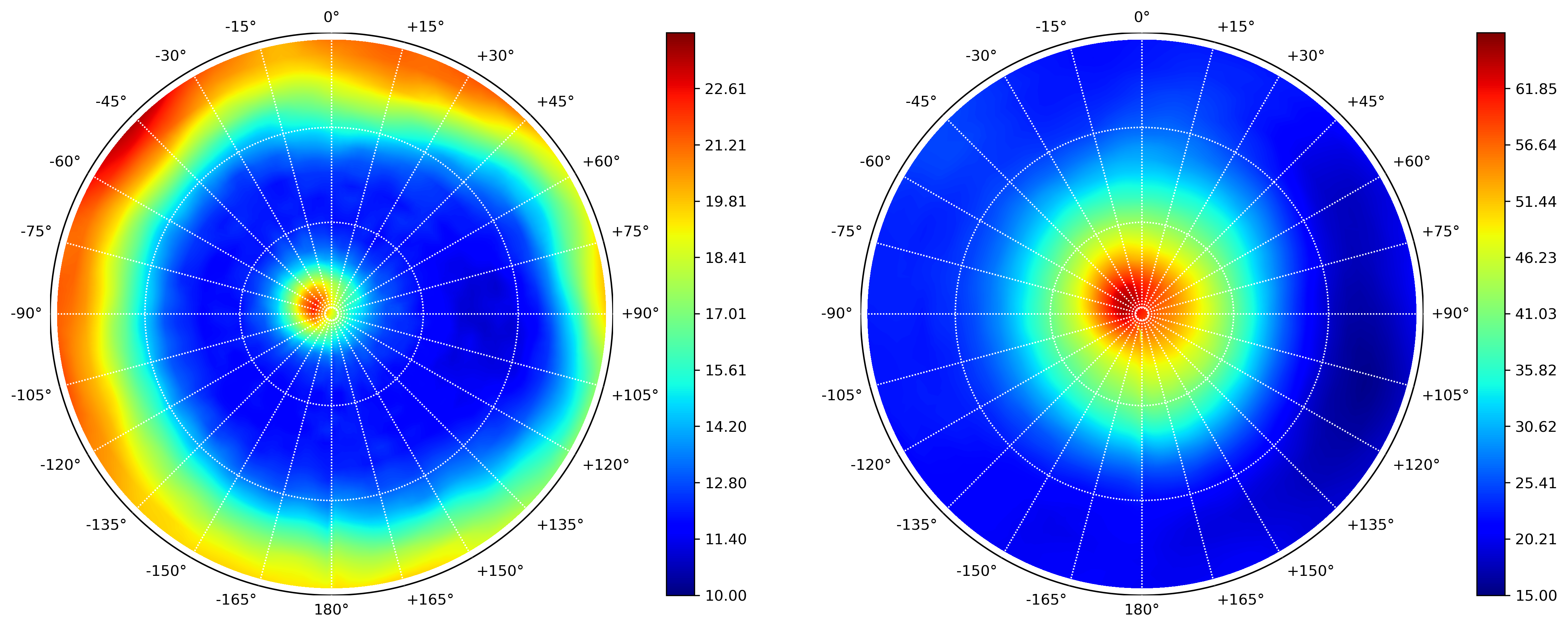}
\caption{South polar stereo-graphic projection of thermal neutron counting rates in the summer (left, $L_s = 140^{\circ} - 180^{\circ}$) and in the winter (right, $L_s = 330^{\circ} - 360^{\circ}$).}
\label{fig:result_fig2compare}
\end{figure}

\begin{figure}[h]
\centering
\includegraphics[width=1\textwidth]{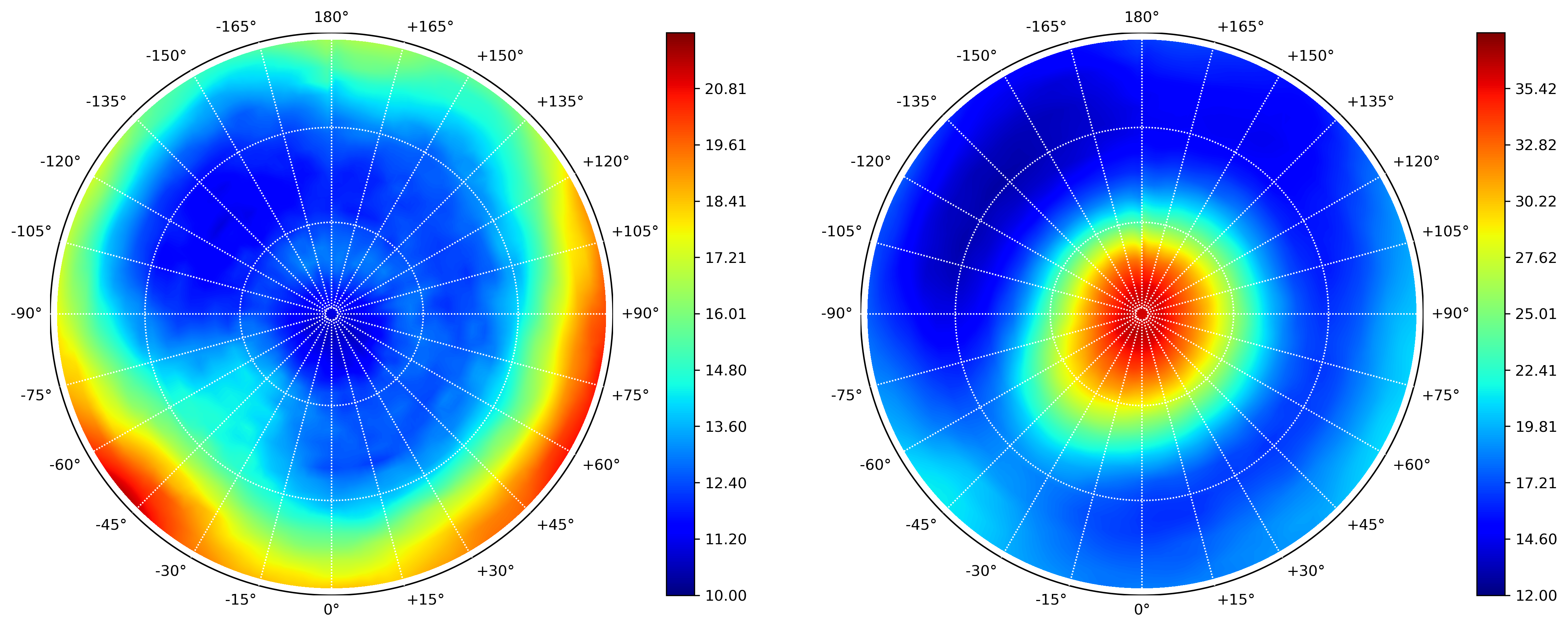}
\caption{North polar stereo-graphic projection of thermal neutron counting rates in the summer (left,$L_s = 110^{\circ} - 150^{\circ}$) and in the winter (right, $L_s = 330^{\circ} - 360^{\circ}$), $0^{\circ} - 20^{\circ}$.}
\label{fig:result_fig2compare_north}
\end{figure}

Stereo-graphic polar projections of the averaged thermal neutron counting rate (Prism 2 - Prism 4) are shown in Figs.~\ref{fig:result_fig2compare} and ~\ref{fig:result_fig2compare_north} for the South and North poles, respectively.  
Each plot shows the thermal neutron counting rate during local summer (left) and local winter (right).  The maps extend to a latitude of 45$^{\circ}$ in the respective hemisphere and are smoothed with a Gaussian filter with 5$^{\circ}$ FWHM to remove random variations between pixels.  The South pole plots in Fig.~\ref{fig:result_fig2compare} use cuts of $L_s = 330^{\circ} - 360^{\circ}$ to define Southern summer and $L_s = 140^{\circ} - 180^{\circ}$ to define Southern winter, similar to the plots in Fig.~2 of \cite{Prettyman2004}.  The North pole plots in Fig.~\ref{fig:result_fig2compare_north} use cuts of $L_s = 110^{\circ} - 150^{\circ}$ to define Northern summer and $L_s = 330^{\circ} - 360^{\circ}, 0^{\circ} - 20^{\circ},$ to define Northern winter.  The summer plots for both poles can also be compared to Fig.~25 in \cite{Maurice2011}, which shows frost free polar maps of thermal neutron counting rates down to $\pm60^{\circ}$.  In the winter, the seasonal caps are clearly identified; in the Southern hemisphere, the peak in neutron counting rate is slightly offset from center towards the Northwest in the plot.  In the summer, the South pole exhibits the perennial CO$_2$ cap that is offset to the West-Northwest in the plot.  Outside of this cap, the counting rates poleward of $-60^{\circ}$ are generally much lower than the lower latitude terrain.  This is similar in the summer at the North pole, and small enhancements in the counting rate above $60^{\circ}$N are similar to observations in \cite{Maurice2011}.

This dataset is intended to be used for studying seasonal effects and comparison of inter-annual variability, therefore the typical data product will be a count rate binned not only in latitude and longitude, but also in year and seasonal $L_s$.  Therefore, the typical uncertainty in each bin, defined as the standard deviation of the data divided by the square-root of the number of entries within each bin, will be larger than \cite{Maurice2011} which focused on removing frost effects to produce global time-averaged count rate maps.  To limit the uncertainties in a given bin to less than 10\%, a limit of 10$^{\circ}$ binning in $L_s$ and 4$^{\circ}$ binning in latitude and longitude is required.  The uncertainty over all years for this binning is shown in Fig.~\ref{fig:result_fig19compare}.  The average uncertainty (solid lines) vary slightly with latitude, due to there being more spread in the data due to frost effects, and turn up at the highest point because no data exist poleward of $\pm$87$^{\circ}$.  The shaded region represents the extent of the inner 80\% of the data.  
\begin{figure}
\centering
\includegraphics[width=1\columnwidth]{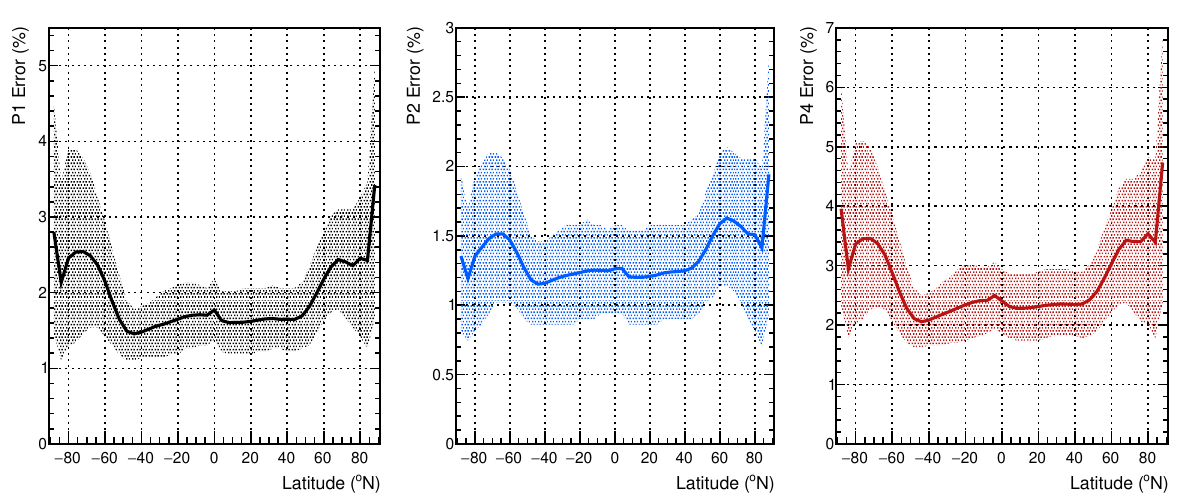}
\caption{Uncertainty in count rate distribution as a function of latitude in 4$^{\circ}$ bins for Prism 1 (left), Prism 2 (middle), and Prism 4 (right).  The average uncertainty (solid lines) is the average over uncertainties in all bins at that latitude (10$^{\circ}$ in $L_s$ and 4$^{\circ}$ in longitude).  The shaded regions correspond to the inner 80\% of all uncertainty values.}
\label{fig:result_fig19compare}
\end{figure}
Similar plots can be made for frost free data only (like Figs.~19 and 26 in \cite{Maurice2011}), and the errors in the polar regions will increase slightly due to a reduction in the number of data entries available, but the spread in the uncertainty will decrease.  Assuming the data is evenly spread over the 8 Martian Years, the errors shown in Fig.~\ref{fig:result_fig19compare} will increase by a factor of 2.8 ($\sqrt{8}$) when considering a single year, leading to errors of $\sim$4\%--7\% for Prism 1, $\sim$3.5\%--5\% for Prism 2, and $\sim$5.5\%--10\% for Prism 4.

\begin{figure}[h]
\centering
\includegraphics[width=0.48\columnwidth]{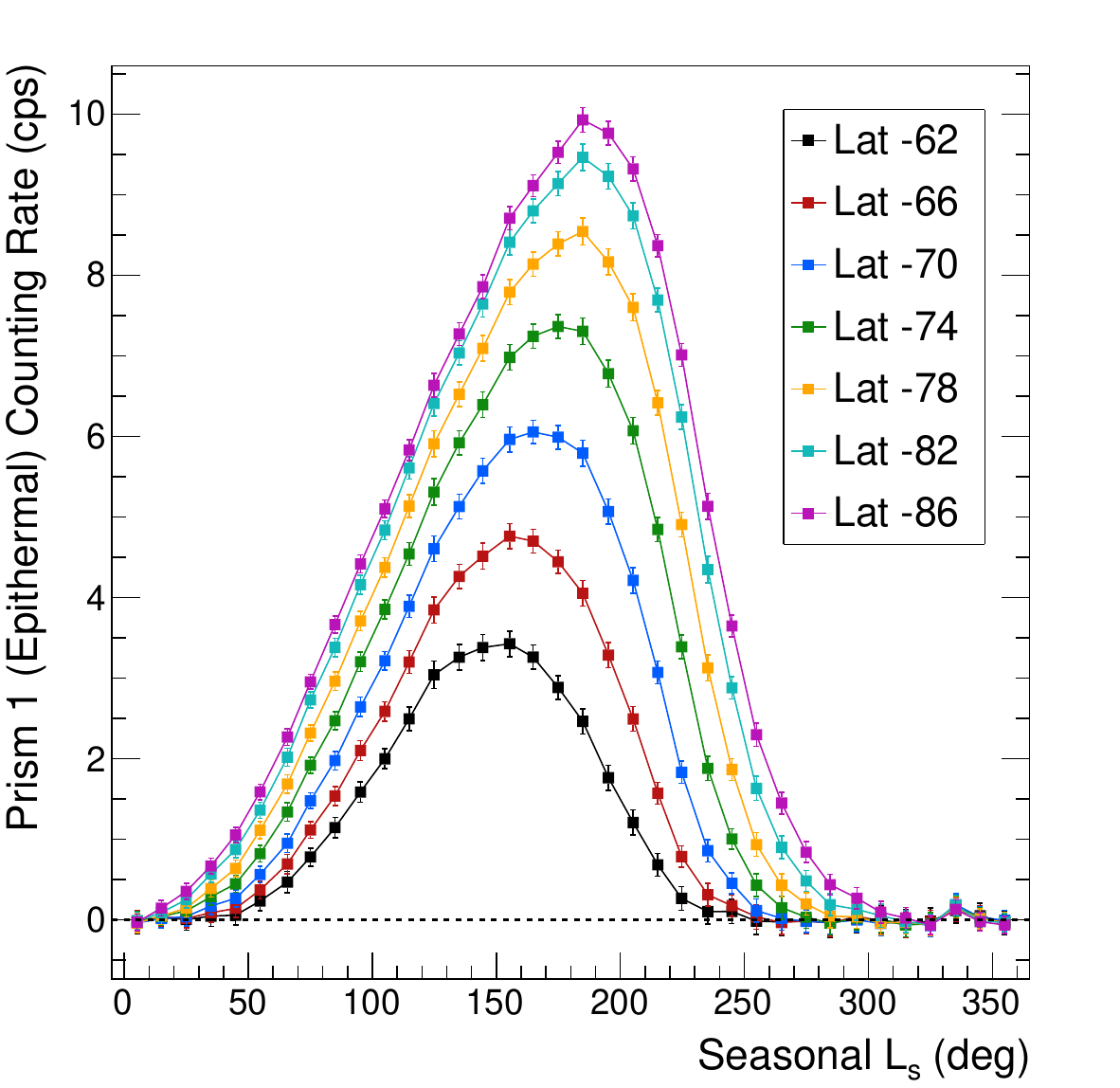}
\includegraphics[width=0.48\columnwidth]{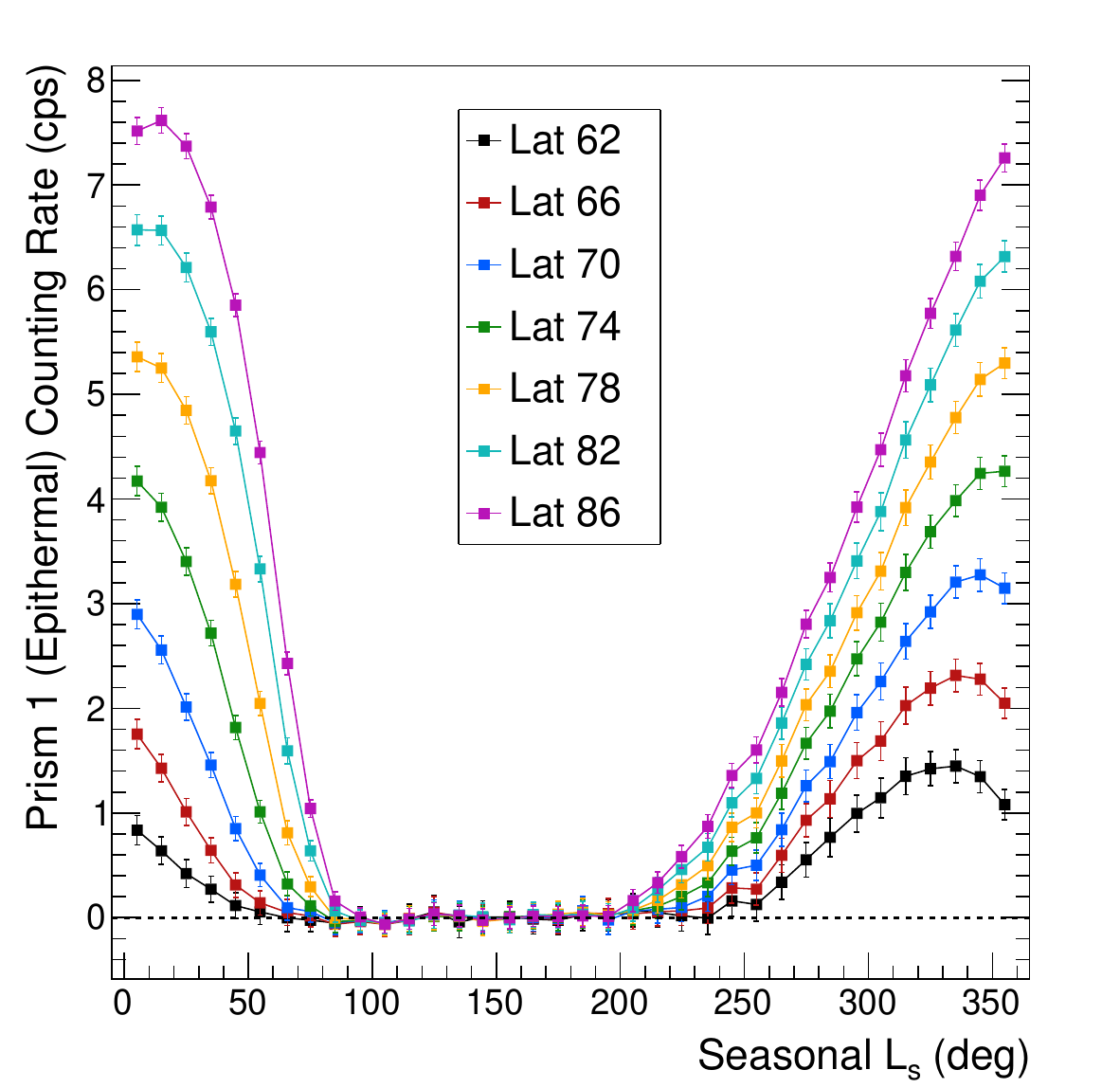}
\caption{(Color online) Example of background-subtracted counting rates for Prism 1 (epithermal) as a function of $L_s$ at the South pole (left) and the North pole (right), averaged over all MY.}
\label{fig:result_p1allyears}
\end{figure}

\begin{figure}[h!]
\centering
\includegraphics[width=0.48\columnwidth]{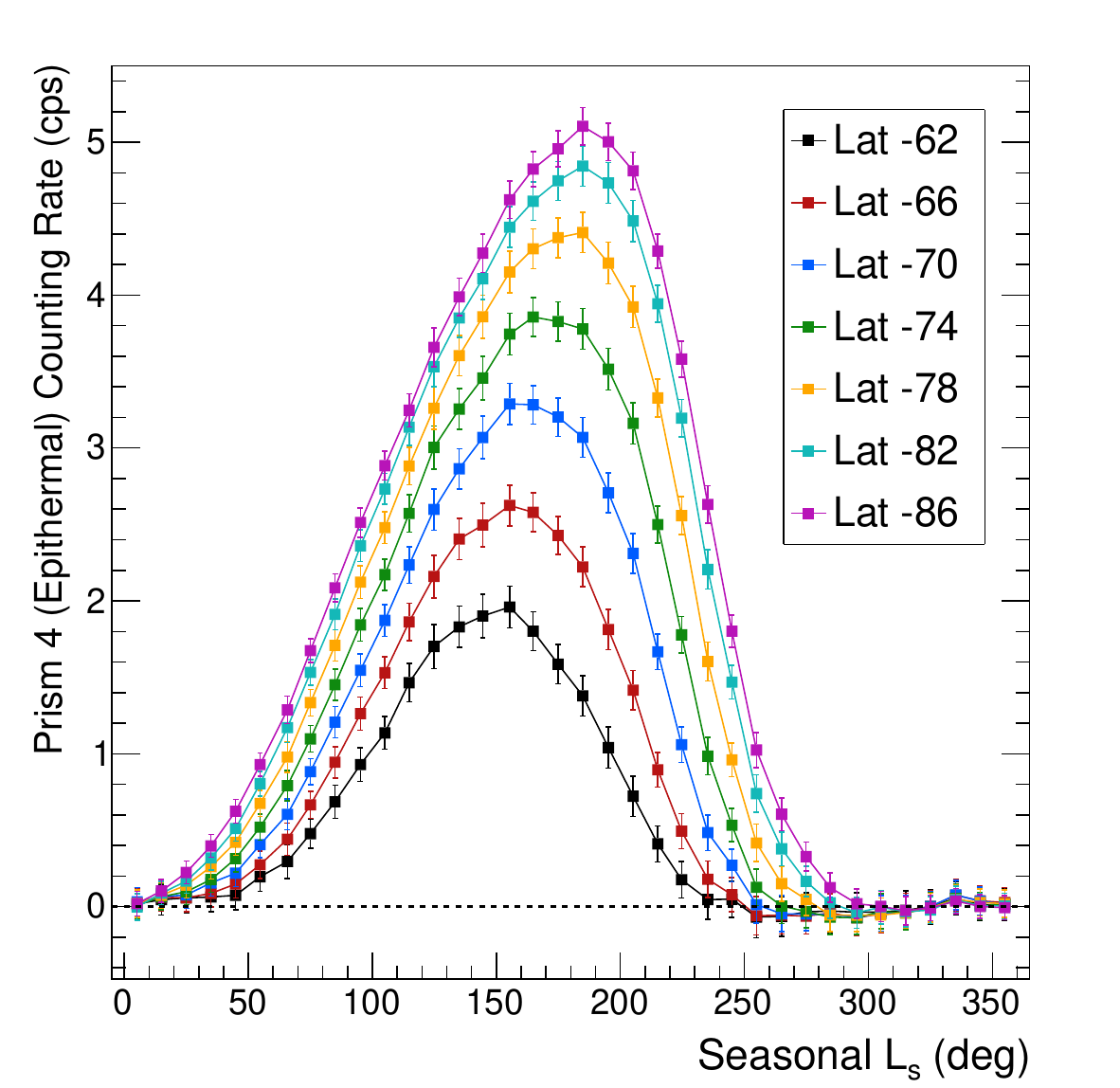}
\includegraphics[width=0.48\columnwidth]{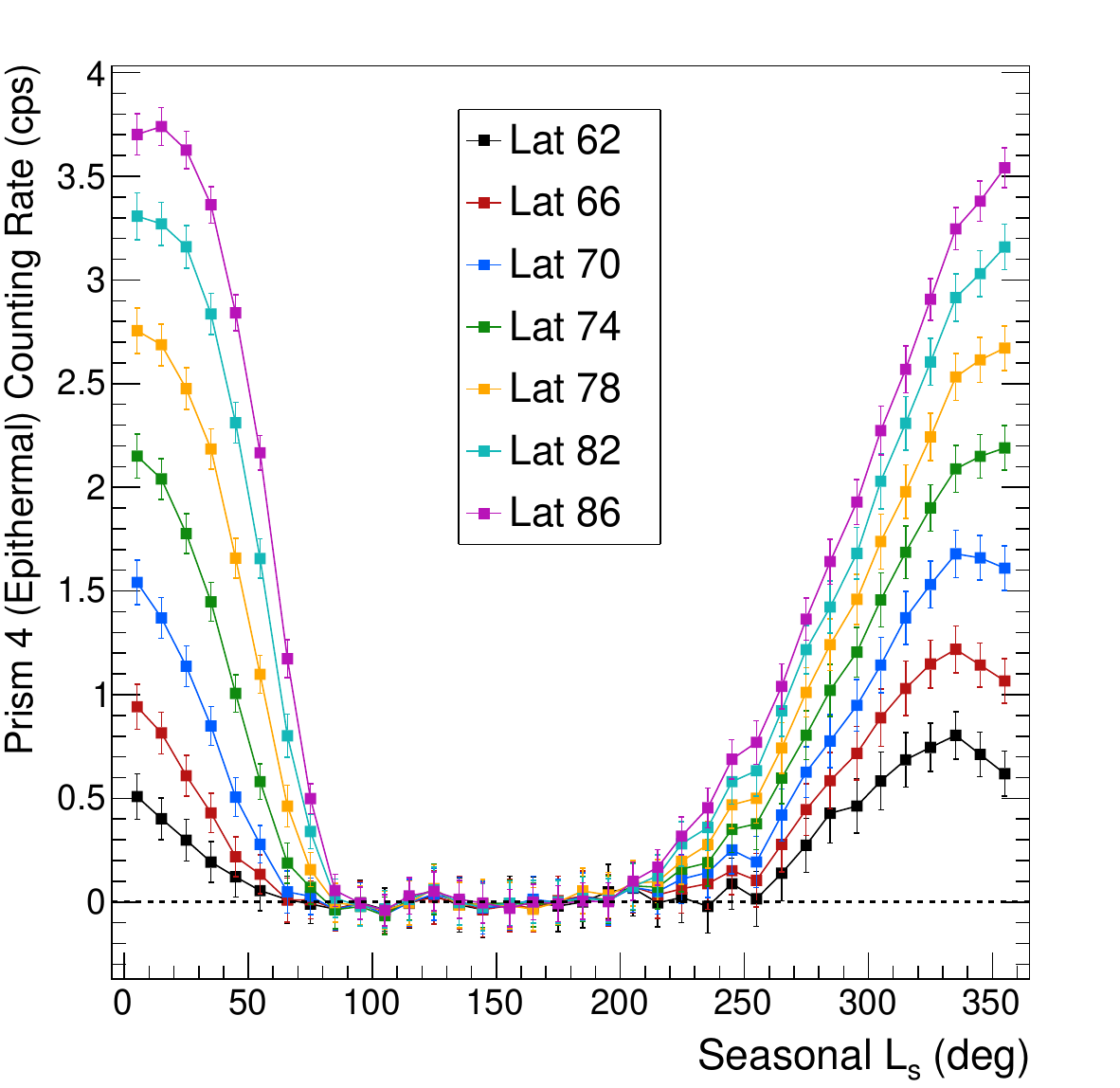}
\caption{(Color online) Same as Fig.~\ref{fig:result_p1allyears} but for Prism 4 (alternate epithermal).}
\label{fig:result_p4allyears}
\end{figure}

\begin{figure}[h!]
\centering
\includegraphics[width=0.48\columnwidth]{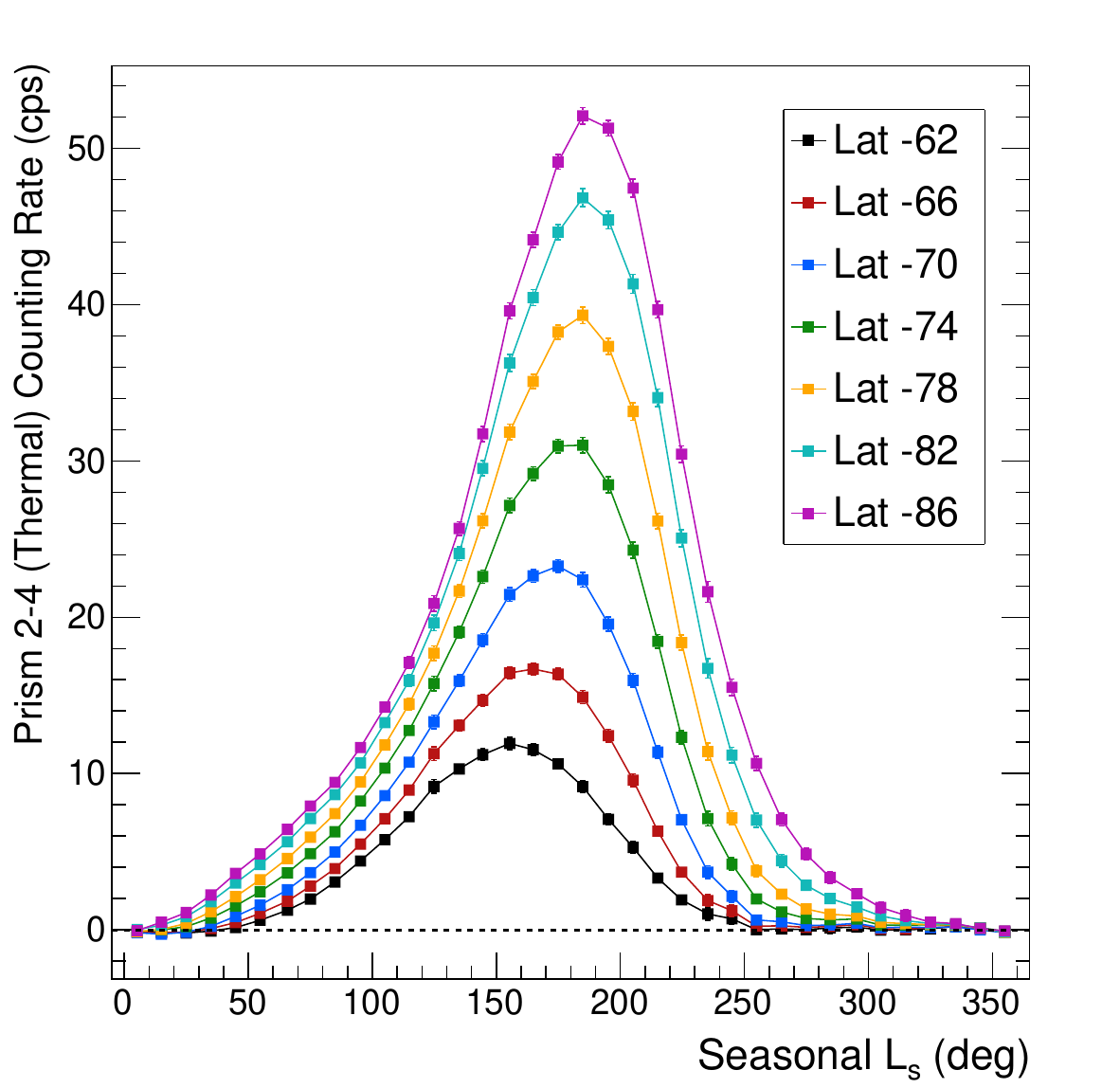}
\includegraphics[width=0.48\columnwidth]{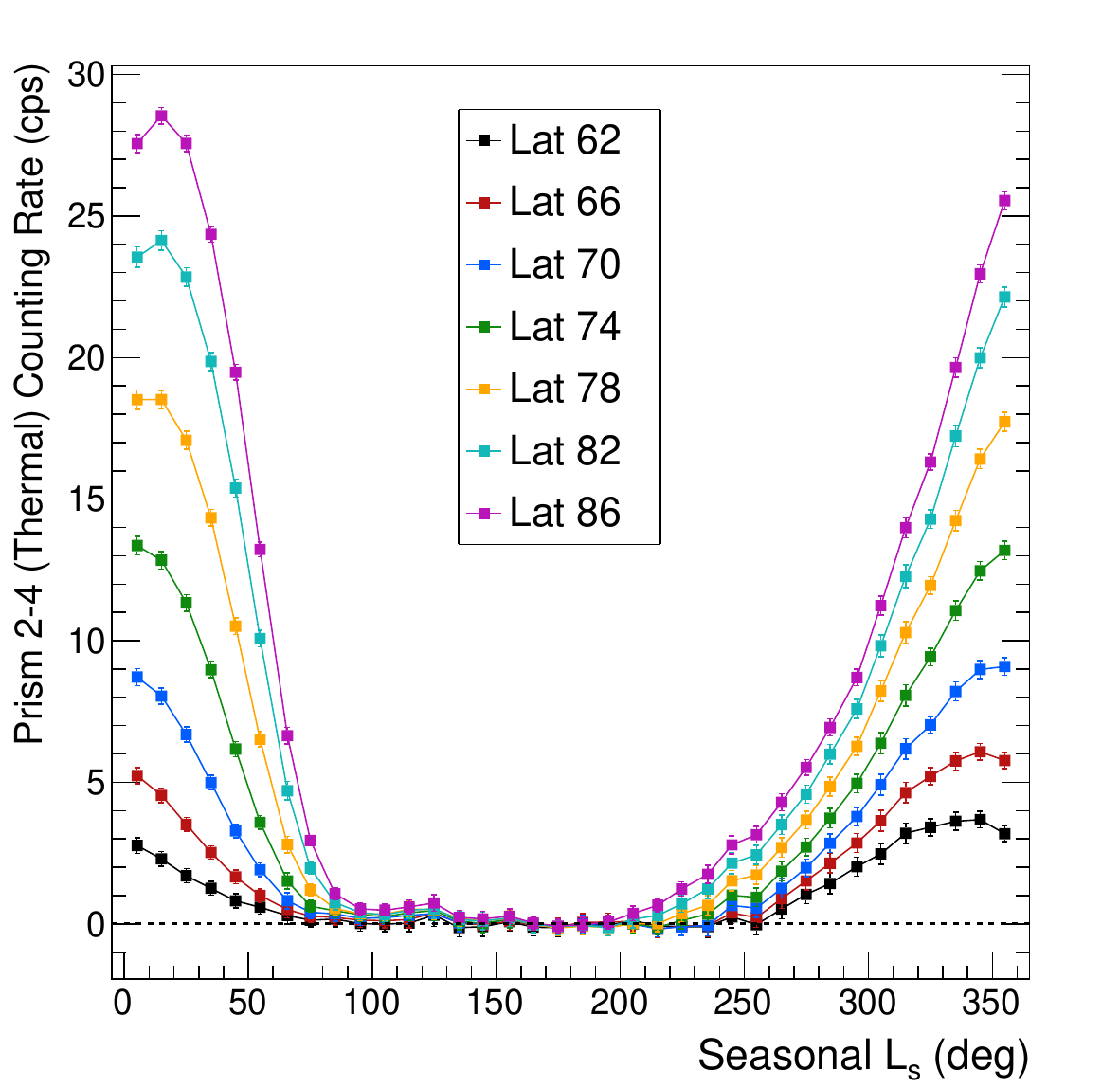}
\caption{(Color online) Same as Fig.~\ref{fig:result_p1allyears} but for Prism 2 - Prism 4 (thermal).}
\label{fig:result_p2m4allyears}
\end{figure}

We generated a map with 10$^{\circ}$ binning in $L_s$ and 4$^{\circ}$ binning in latitude and longitude and plotted counting rate trends as a function of $L_s$ to perform a preliminary comparison of inter-annual variability.  The counting rate averaged over all Mars years versus $L_s$ for latitude bins in the polar regions are shown in Fig.~\ref{fig:result_p1allyears}, \ref{fig:result_p4allyears}, and \ref{fig:result_p2m4allyears}, for Prism 1 (epithermal neutrons), Prism 4 (alternate epithermal neutrons), and Prism 2 - Prism 4 (thermal neutrons), respectively.  The counts are integrated over all longitude and the latitude bin noted in the legend is the center point within the 4$^{\circ}$ bin.  The counting rates have a rough summer-time background subtraction applied by using the average of counting rates during summer so that the counts above the summer-time baseline counting rate can be compared across latitudes.  Both epithermal analogs have similar trends, with the Prism 4 counting rates about a factor of 2 lower than the Prism 1 counting rates.  The thermal counting rates have a much different trend and much higher counting rate.  At this stage, atmospheric effects have not been removed from the data.

\begin{figure}[h!]
\centering
\includegraphics[width=1\columnwidth]{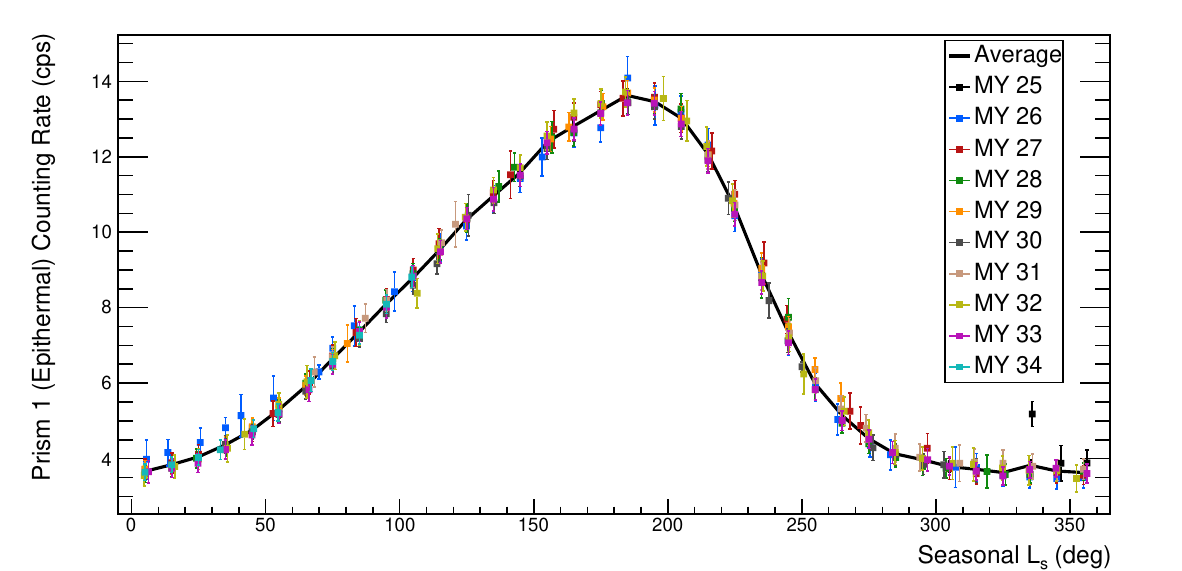}
\includegraphics[width=1\columnwidth]{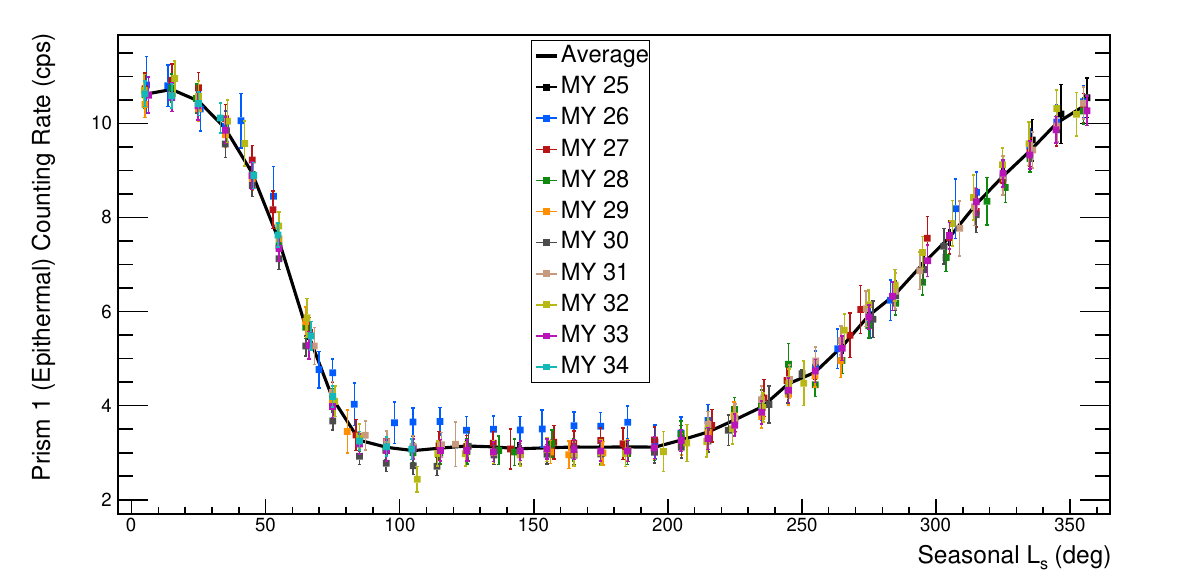}
\caption{(Color online) Prism 1 (epithermal) counting rates for 86$^{\circ}$S (top) and 86$^{\circ}$N (bottom) as a function of $L_s$ separated by MY.}
\label{fig:result_p1_sepyears}
\end{figure}

\begin{figure}[h!]
\centering
\includegraphics[width=1\columnwidth]{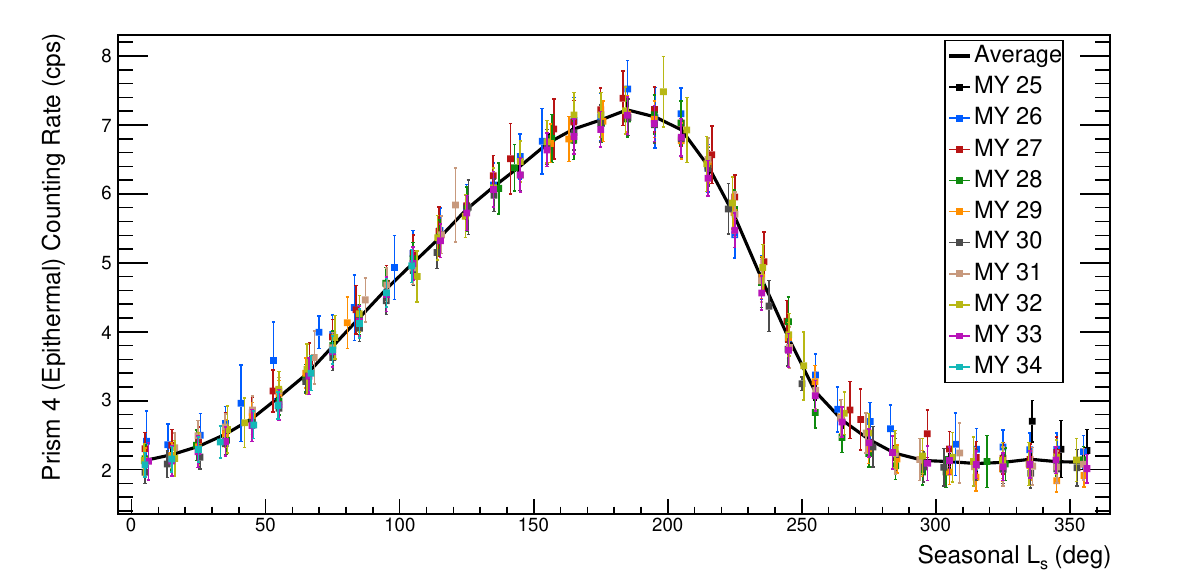}
\includegraphics[width=1\columnwidth]{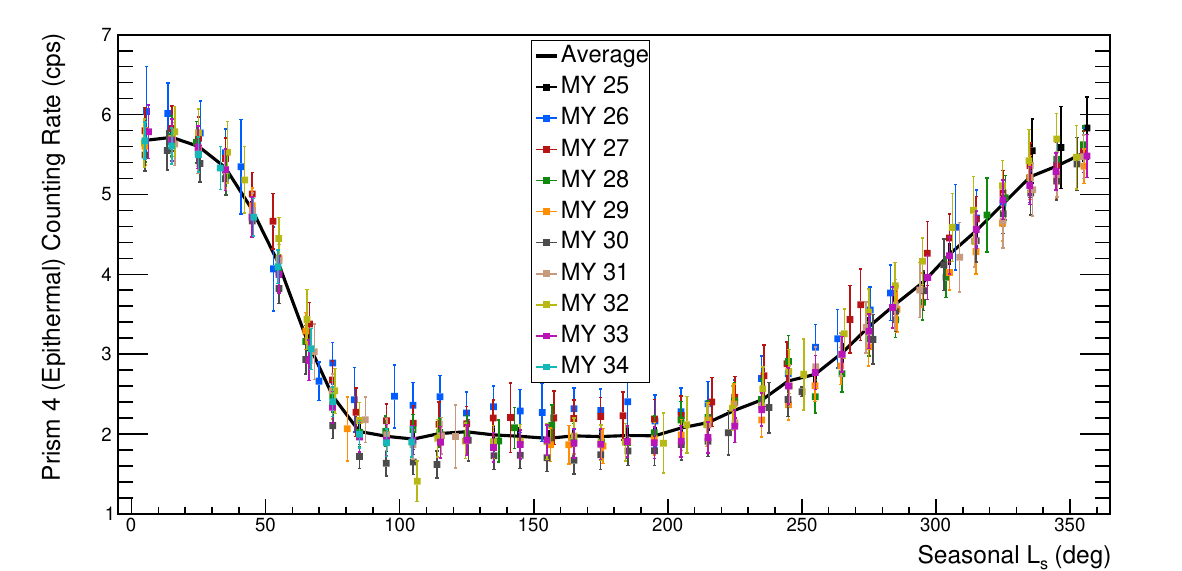}
\caption{(Color online) Same as Fig.~\ref{fig:result_p1_sepyears} but for Prism 4 (alternate epithermal).}
\label{fig:result_p4_sepyears}
\end{figure}

\begin{figure}[h!]
\centering
\includegraphics[width=1\columnwidth]{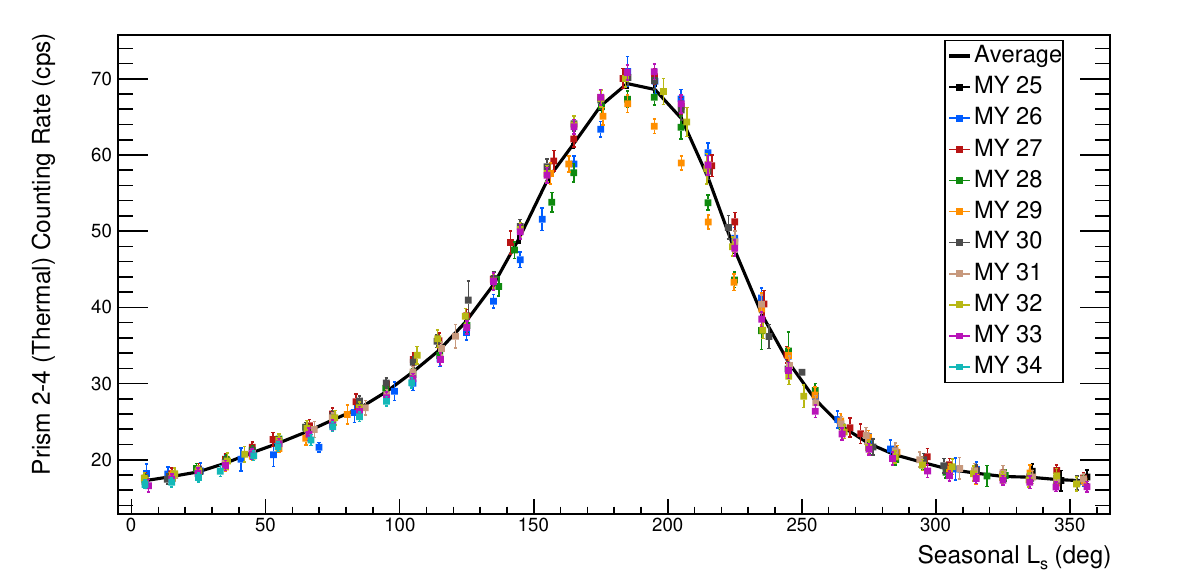}
\includegraphics[width=1\columnwidth]{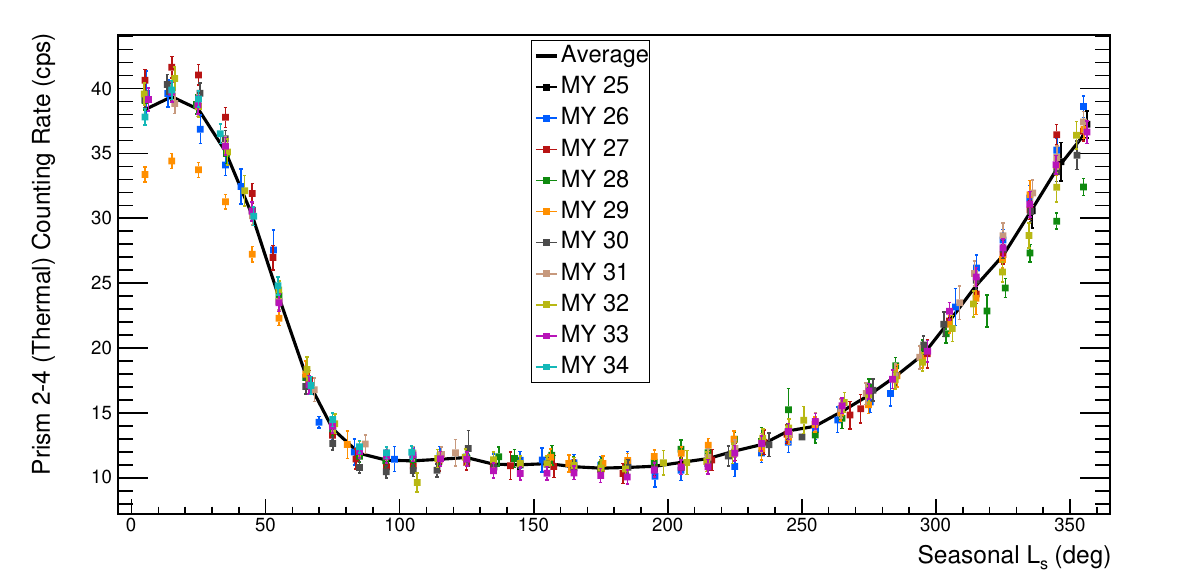}
\caption{(Color online) Same as Fig.~\ref{fig:result_p1_sepyears} but for Prism 2 - Prism 4 (thermal).}
\label{fig:result_p2m4_sepyears}
\end{figure}

The counting rates as a function of $L_s$ are separated by MY for the 86$^{\circ}$ latitude band in Figs.~\ref{fig:result_p1_sepyears}, \ref{fig:result_p4_sepyears}, and \ref{fig:result_p2m4_sepyears} for Prism 1 (epithermal neutrons), Prism 4 (alternate epithermal neutrons), and Prism 2 - Prism 4 (thermal neutrons), respectively.  
These plots also show the counting rate averaged over all MY for reference as a solid line.  In general, the counting rate trends from year to year are very reproducible and agree within the uncertainties of the data.  The notable exception to the reproducibility, which was previously mentioned, is the thermal neutron counting rate from $L_s\sim315^{\circ}$ in MY 28 through $L_s\sim45^{\circ}$ in MY 29.  At the peak of Northern winter, the MY 29 counting rates are about 14\% lower than the average of the other Mars years.  This drop in counting rate, which is suggestive of less CO$_2$ deposition, occurs after the planet-wide global dust storm that emerged around $L_s \sim 265^{\circ}$ in MY 28 \cite{Smith2009,Wang2015}.  Observations in the Mars Climate Sounder thermal infrared data \cite{Piqueux2015b} show the Northern seasonal cap is $\sim$10\% smaller in spatial extent following the MY 28 global dust storm.  In future work, it will be interesting to process and analyze the MONS data through 2018, as another global dust storm with an onset in $L_s = 185^{\circ}$ occurred in MY 34.

\section{Summary \& Future Work}

In summary, we have performed a full re-analysis of Mars Odyssey Neutron Spectrometer data, extending the coverage of MONS data through the end of 2017 to cover 8 Martian Years.  This paper summarizes the data analysis procedure, including data reduction and data corrections, to document and provide the necessary understanding to process raw MONS data from the PDS and utilize this new dataset.  Example results based on this new dataset include frost-free global counting rate maps and maps of the polar regions and were presented in Section~\ref{sec:results}.  These maps were qualitatively compared to previous analyses of the MONS data performed in \cite{Prettyman2004,Maurice2011} and found to show similar trends.  Due to choices made in processing of this dataset, the overall normalization of the data is different than the previous analyses.  We showed the averaged counting rates for different latitude bands in the polar regions, which show the typical latitude dependence of the counting rates as a function of $L_s$.  Preliminary results on inter-annual variability of the seasons CO$_2$ caps were also presented in Section~\ref{sec:results}, which show reproducibility in the Southern seasonal cap based on both thermal and epithermal neutron counting rates, and reproducibility in the Northern seasonal cap based on the epithermal neutron counting rates and in thermal counting rates in all years except MY 28 - MY 29, following a global dust event.  The variation in the thermal neutron counting rates in MY 28 - MY 29 may be due to a combination of cap properties and atmospheric properties, and will be explored in future work that will include an inter-annual comparison across all the latitude bins.

Work utilizing this new dataset is ongoing by this team.  We are currently performing the necessary simulation and modeling efforts to normalize the data and convert counting rates to CO$_2$ frost thickness in both the North and South polar regions and better understand atmospheric effects.  With these efforts, we will be better able to interpret the overall properties of the seasonal caps and how the MY 28 global dust storm impacted the overall mass of CO$_2$ deposited and any changes in the extent in the Northern seasonal cap following this event.   This work will be the subject of future papers.

\section{Data Availability}
The unbinned time-series dataset related to this article can be found at \url{https://doi.org/10.5281/zenodo.3267970}, hosted at Zenodo \cite{Mesickdata}.  The neutron counting rate maps binned by mars year, mars solar longitude, latitude, and longitude, will be published in the Planetary Data System Geosciences node in 2020 at the conclusion of funded NASA work.  Questions related to the use of the data can be directed to the corresponding author.

\section*{Acknowledgments}
Research presented in this paper was supported by the Laboratory Directed Research and Development program of Los Alamos National Laboratory under project number 20160672PRD3 and the NASA Mars Data Analysis Program under project number 80HQTR170004.  This research used resources provided under the Los Alamos National Laboratory Institutional Computing Program, which is supported by the U.S. Department of Energy National Nuclear Security Administration under Contract No. 8923328CNA000001.  The authors would like to acknowledge the reviewers for their helpful comments in improving the manuscript.

\newpage
\section*{Appendix A: Simulation Tools}

\subsection*{Neutron Flux Signal}

The radiation transport simulation package Geant4 \cite{Agostinelli2003} was utilized to develop a tool to simulate the expected neutron leakage from Mars at the top of the atmosphere.  Benchmarking of this simulation package \cite{Mesick2018} has shown that with the appropriate choice of physics model, Geant4 and MCNP6.2 \cite{MCNP6.2} produce similar results but both slightly over-predict neutron density profiles measured in the Lunar Neutron Probe Experiment \cite{Woolum1975}.  The shapes of the curves generally match the data well, and therefore with proper normalization these types of simulation tools are helpful in the interpretation of planetary neutron data.

For the simulations relevant to determining the change in neutron flux with atmospheric density, a single-layer model with varying amounts of water-equivalent hydrogen (WEH) and varying atmosphere thicknesses were performed.  The soil composition was based on the S21 average composition from \cite{Diez2008}, with elemental compositions given in Table~\ref{table:s21}.  A fixed Cl abundance of 0.517\% was added to this composition, close to the average as measured by the Mars Odyssey gamma-ray instrument.  The size of the atmosphere normalization correction to the counting rate depends on the average WEH of the soil, and published maps of average WEH from the Mars Odyssey gamma-ray detector that are available in the PDS were used to estimate this within the belly band region.  After adding in Cl abundance and the appropriate amount of H$_2$O, the S21 elemental abundances were scaled uniformly so that the total elemental abundance summed to unity.  The density of the soil was assumed to be 1.8~g/cm$^3$.
\begin{table}[h]
\centering
\caption{Elemental weight fractions in the base soil composition based on S21 from \cite{Diez2008}}
\label{table:s21}
\begin{tabular}{|c|c|c|c|c|}
\hline
O & 0.4223 & & K & 0.0022 \\
\hline
Na & 0.0147 & & Ca & 0.0419 \\
\hline
Mg & 0.0478 & & Ti & 0.0052 \\
\hline
Al & 0.0464 & & Cr & 0.0016 \\
\hline
Si & 0.2075 & & Mn & 0.0033 \\
\hline
S & 0.0278 & & Fe & 0.1466 \\
\hline
\end{tabular}
\end{table}

The composition of the atmosphere in the simulation was based primarily on values used in \cite{Prettyman2004}, which come from the Viking data \cite{Lewis1984} with minor modifications.  The concentrations of CO$_2$, H$_2$O, N$_2$, and Ar from \cite{Prettyman2004} are 96.93\%, 0.054\%, 2.7\%, and 1.6\%, respectively.  In addition, an O$_2$ concentration of 0.13\% was assumed, taken from \cite{Lewis1984}. The atmosphere was simulated out to 40~km above the surface, with 20 layers of exponentially decreasing density (exp($-h/H$), where $H$ is the scale height) and a scale height of 10.8~km, similar to the layering used in simulations by \cite{Jun2013}.

As described in \cite{Feldman1989}, gravitational binding of neutrons can effect the measured flux spectra.  On Mars, the gravitational binding energy is 0.132~eV.  Neutrons below this energy can return to the surface and re-interact.  Gravitational binding of neutrons was implemented in our simulation by a reflecting boundary at the top of the atmosphere that reflected neutrons below the binding energy back to the surface.  Since the surface return time $\Delta$t (derived in \cite{Feldman1989}) can be on the order of the neutron decay lifetime, a weighting factor exp($-\Delta t/\tau$) was applied based on the probability of neutron decay assuming the most recent value of the neutron free lifetime, $\tau$ = 880.2~s \cite{PDG2018}.  The epithermal and fast neutron flux are not affected by gravitational binding, however, the thermal neutron flux is almost 50\% higher at the top of the atmosphere when gravitational binding is included.

Simulations were run for WEH values of 1\%, 3\%, 4\%, 5\%, 6\%, and 8\% (spanning the range of measured values in the belly band region) with total atmospheric thicknesses ($\rho_a$) of 4, 8, 12, 16, 20, and 24~g/cm$^2$ (spanning the range of values predicted by the GCM model \cite{Forget1999} in the belly band region) at each WEH point.  In these simulations, the GCR flux was modeled following the Castognali \& Lal model \cite{Munoz1975} with a solar modulation of $\phi = 900$~MV.  More details about this model and other GCR models can be found in \cite{Mesick2018}. At this stage, the neutron energy and angle at the top of the atmosphere are recorded.  For the atmospheric correction since only ratios are being compared, the absolute GCR flux is not important.

\subsection*{Detector Response}

\begin{figure}[h!]
\centering
\includegraphics[width=0.65\textwidth]{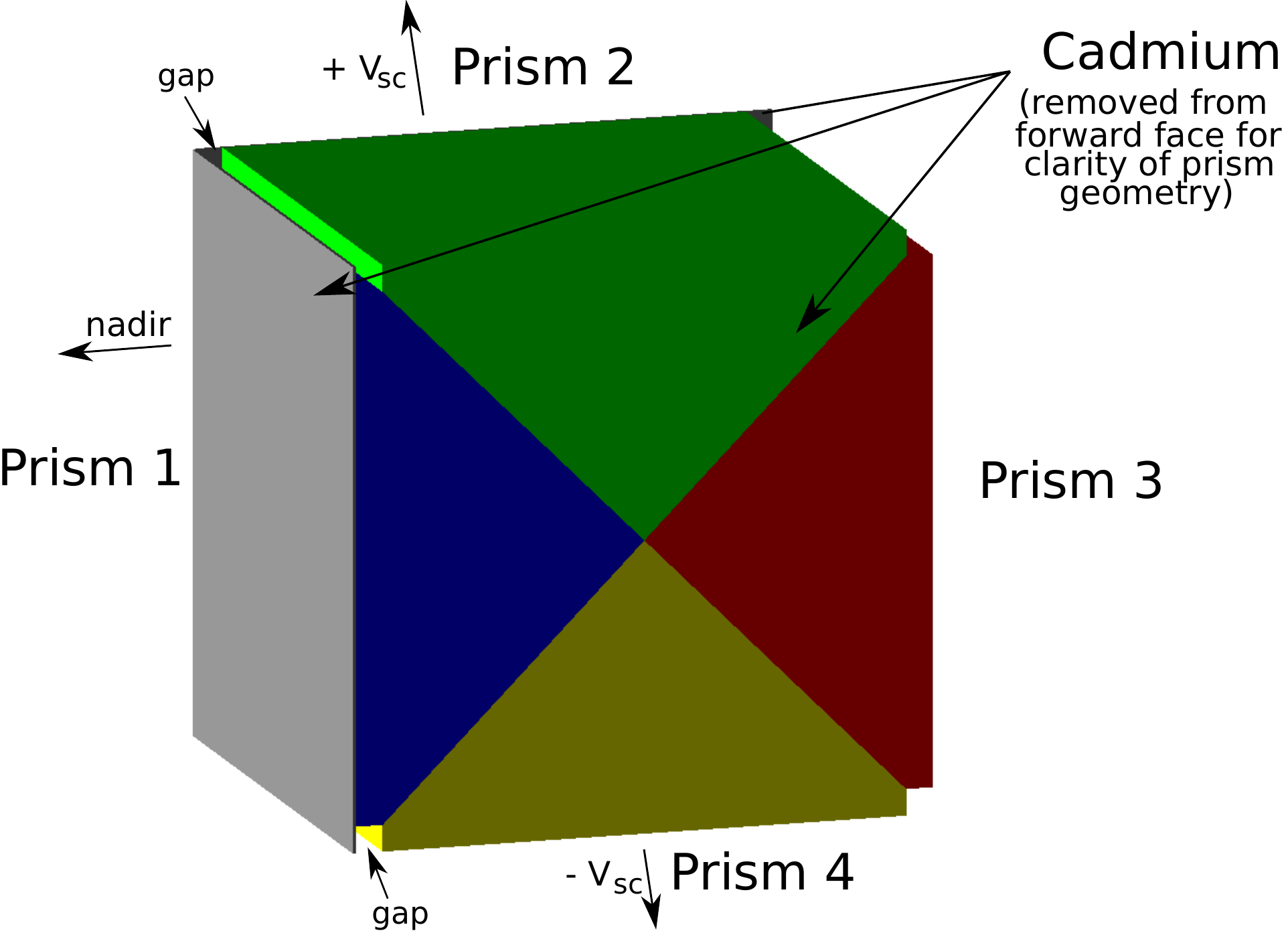}
\caption{Geometry modeled in the detector response simulations.  The gaps in the cadmium covering are indicated.  The cadmium covering the front face of the view (Prism side) was removed for clarity to view the Prism geometry.}
\label{fig:mons_sim}
\end{figure}

The MONS instrument response was modeled in a separate simulation utilizing Geant4; the geometry is shown in Fig.~\ref{fig:mons_sim}.  The geometry of the four boron-loaded plastic prisms was modeled, including cadmium covering where appropriate.  The gaps between the Prism 1 face and the cadmium sheet where thermal neutrons can leak in was also modeled.  The external spacecraft was not included in the model based on results from \cite{Prettyman2004}, which describes that Prism 1 is well shielded from spacecraft background, and that Prism 2 and Prism 4 have the same response so that subtraction of Prism2 - Prism 4 to determine the thermal neutron counting rate will effectively remove the spacecraft background.

\begin{figure}[t]
\centering
\includegraphics[width=0.9\textwidth]{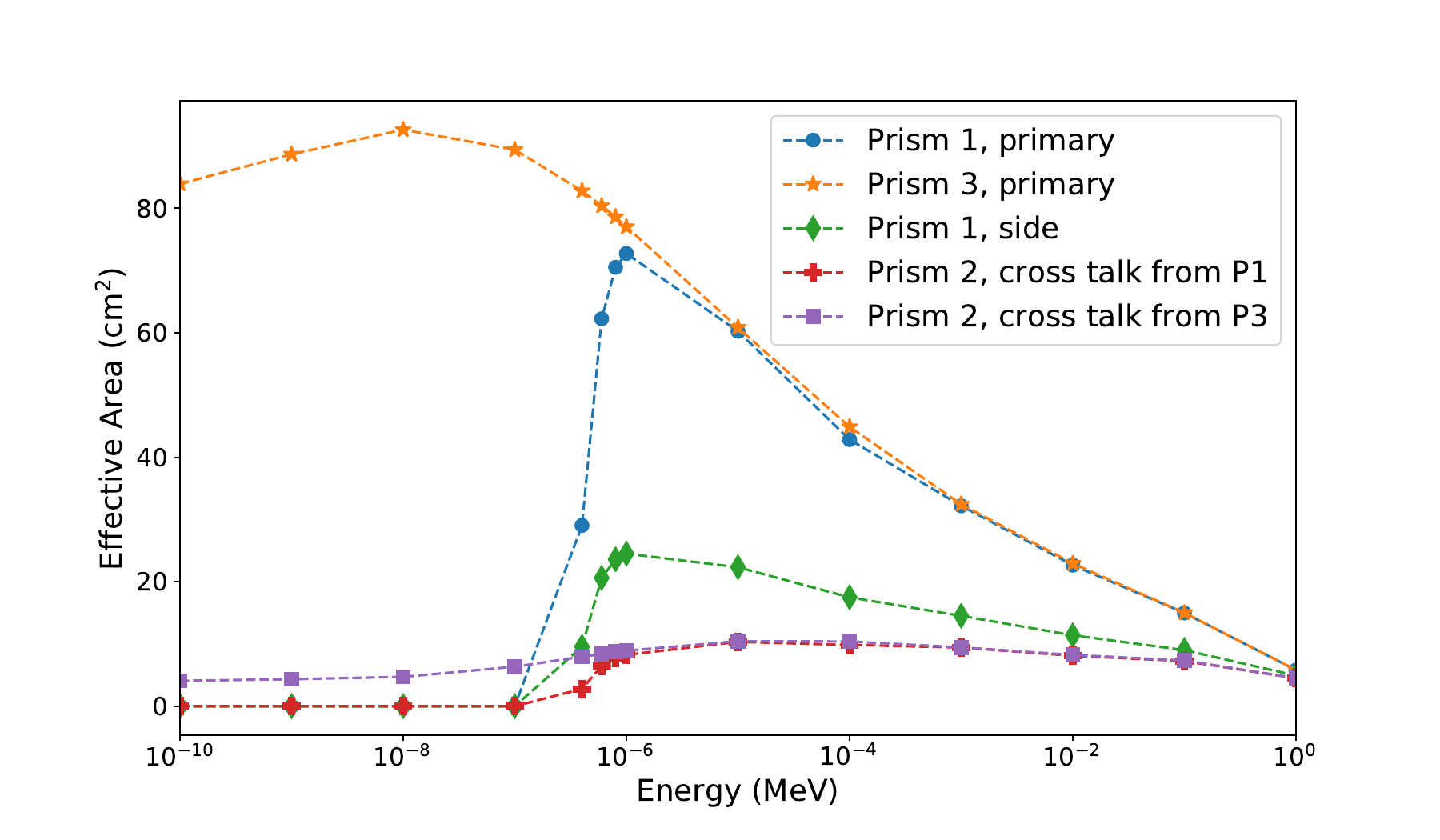}
\caption{Effective area for neutron capture as a function of energy, for an incident angle of 0$^{\circ}$, see text for details.}
\label{fig:mons_ea}
\end{figure}
The efficiency for neutron capture was simulated with an $11\times11$~cm$^2$ planar source placed directly on top of a given prism face, which fully covers the active area including the gaps.  Neutrons were uniformly generated from random points within the rectangular dimensions.  Neutron energies of 1$\times$10$^{-10}$ MeV and increasing in increments of one decade to 1 MeV were simulated, and at each energy the neutrons originated at angles from 0 to 85 degrees in increments of 5$^{\circ}$.  A simulation with the source incident on Prism 1 provides the efficiency for epithermal neutron detection.  A simulation with the source on Prism 3 provides the efficiency for thermal neutron detection, and serves as a proxy for Prism 2 and Prism 4 through an angular transformation.  A simulation with the source on the side of the prisms was also performed.  In addition to the efficiency being tabulated for events hitting the primary prism in each configuration, the efficiency of cross-talk events, \textit{e.g.} events detected in Prism 2, 3, or 4 when the source was on Prism 1, was also tabulated.  Figure~\ref{fig:mons_ea} shows the effective area at 0$^{\circ}$ incident angle for Prism 1 (epithermal neutrons), Prism 3 (thermal neutrons), the side of Prism 1, and examples of cross talk events detected in Prism 2 when the source is on Prism 1 or Prism 3.  The effective area is the simulated efficiency multiplied by the source area, and the total effective area is the summation of the primary + cross talk effective areas.


The simulated MONS counting rate was calculated by taking the simulated neutron current at the top of the Mars atmosphere (40~km) and combining it with the appropriate effective area tables from the Geant4 detector response simulations.  However, several steps in between occur to account for the ballistic trajectory of the particles, the gravitational binding of neutrons, and the spacecraft velocity.  The procedure to calculate the count rate for each prism includes equations derived in \cite{Feldman1989} and follows the steps below.
\begin{enumerate}
\item Transform the energy and angle to account for the ballistic trajectory due to the gravity of Mars, 
\item Transform the energy and angle into the spacecraft motion frame, 
\item Interpolate the effective area of the incident prism from the final energy and angle of the neutron, and
\item Account for the neutron lifetime.
\end{enumerate}

The energy and angle of the simulated neutrons were transported to the spacecraft orbit assuming ballistic trajectories. The final energy of the neutron is \cite{Feldman1989} $K_r$ = $K-V\frac{(R-{R_M})}{R}$, where $K$ is the energy of the neutron leaving the surface of Mars, $V = GMm/R_M$ is the gravitational binding energy of Mars (0.132 eV), $G$ is the gravitational constant, $M$ is the mass of Mars, $m$ is the mass of the neutron, $R_M$ is the radius of Mars, and $R$ is the distance from the center of Mars to the spacecraft orbital altitude.  The neutron incident angle was adjusted for the ballistic trajectory caused by the binding energy of Mars \cite{Feldman1989}: $\mu_r^2=1-\left(\frac{R_M}{R}\right)^2\frac{K}{K_R}(1-\mu^2)$, where $\mu$ is the cosine of $\theta$, the angle the neutron leaves the surface, and $\mu_r$ is the cosine of $\theta_R$, the angle of the neutron at orbit.

The new energy was transformed into the reference frame of the spacecraft depending on which prism the neutron was determined to hit. Using Galilean velocity transformations the constant speed of the spacecraft (3380~m/s) is added (subtracted) to the component of the neutron velocity in the direction of the spacecraft motion if the neutron hit Prism 2 (4).  For Prism 1 and the prism side, a random sampling of azimuthal $\phi$ values were assigned to each neutron angle and energy, and the incident velocity was calculated with the $\phi$ value included.  This was to account for all possible points the neutrons may hit the detector from, and the results were averaged.  This correction affects what energy each Prism ``sees" based on the motion of the spacecraft, which impacts the effective area.  Prism 2 is heading ``toward" the neutrons and will crash into them, adding to their incident energy and Prism 4 is traveling ``away" from the neutrons taking away some of their energy upon impact.  In addition to this effect, the neutron flux itself must also be corrected for the shift in detected neutron energy.  The original neutron flux and velocity at the top of the atmosphere were used to calculate the neutron density. This neutron density was then multiplied by the velocity of the neutron at the spacecraft to get the neutron flux in the reference frame of the moving spacecraft.  Finally, given the definition of angles in the detector response simulations, an angle transformation of 90$^{\circ}$ is applied to events hitting Prism 2, Prism 4, and the prism side.

The effective area tables were interpolated in energy and angle using the Python 2D SciPy interpolation package.  For events hitting Prism 1, the Prism 1 effective area tables were used.  For events hitting Prism 2 or Prism 4, the Prism 3 effective area tables were used.  For events hitting the side of the prism cube, the Prism 1 side effective area tables were used.  The primary signals (events hitting only the primary prism, and ignoring contributions from the side) account for 65\% of Prism 1 and Prism 4 events and 85\% of Prism 2 events.

Finally, a weighting factor is applied to account for the probability of neutron decay during the transport of neutrons from 40~km to the spacecraft orbit, using a neutron lifetime of $\tau$ = 880.2~s \cite{PDG2018} and the neutron transit time \cite{Feldman1989} (note there is a sign error in this equation as found in \cite{Feldman1989}, which has been corrected here):
\begin{eqnarray}
&&\Delta t_r = \frac{R_M(m/2V)^{1/2}}{2\left[1-K/V\right]^{3/2}}\times \\ \nonumber
&&\left\{2\mu\left(1-\frac{K}{V}\right)^{1/2}\left(\frac{K}{V}\right)^{1/2}\left[1-\left(\frac{\tan^2\theta}{\tan^2\theta_R}\right)^{1/2}\right] + \sin^{-1}\left(\frac{B}{[A^2+B^2]^{1/2}}\right) + \sin^{-1}\left(\frac{1-2K_R/V_R}{[A^2+B^2]^{1/2}}\right)\right\}~,
\end{eqnarray}
where $A = \left[4(K/V)(1-K/V)\mu^2\right]^{1/2}$, $B = 2K/V-1$ for $K/V<1$, $m$ is the mass of the neutron, and the other constants have already been defined.

\subsection*{GCR Correction Details}

\begin{figure}[h!]
\centering
\includegraphics[width=0.48\textwidth]{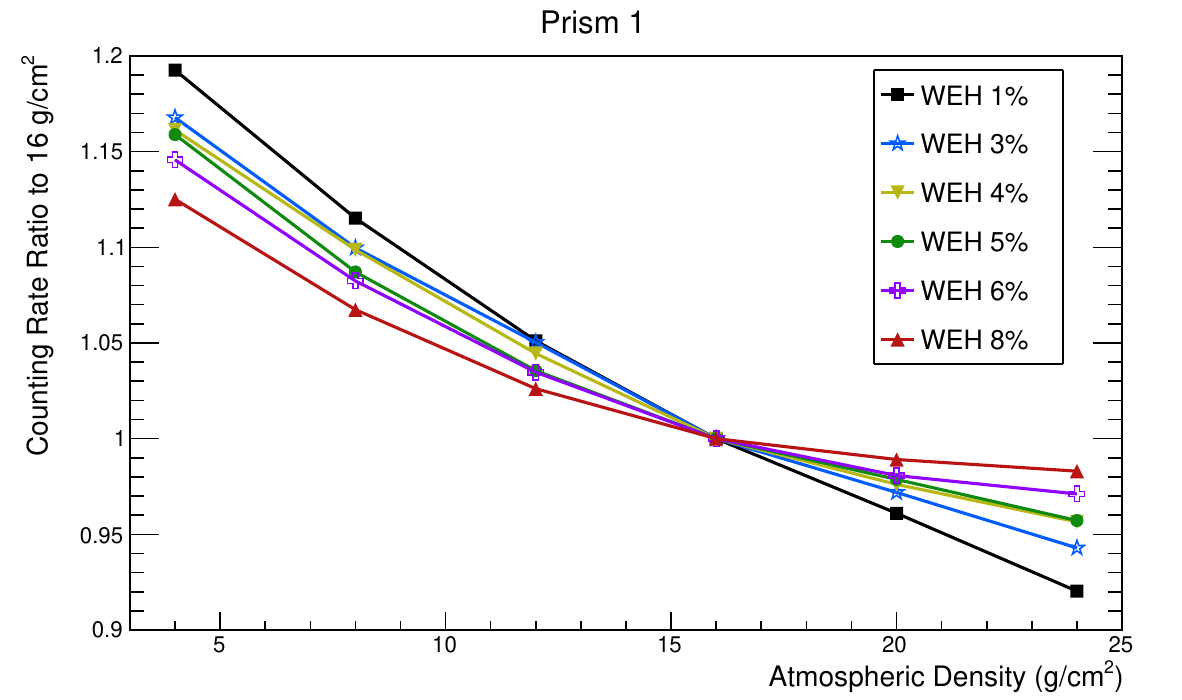}
\includegraphics[width=0.48\textwidth]{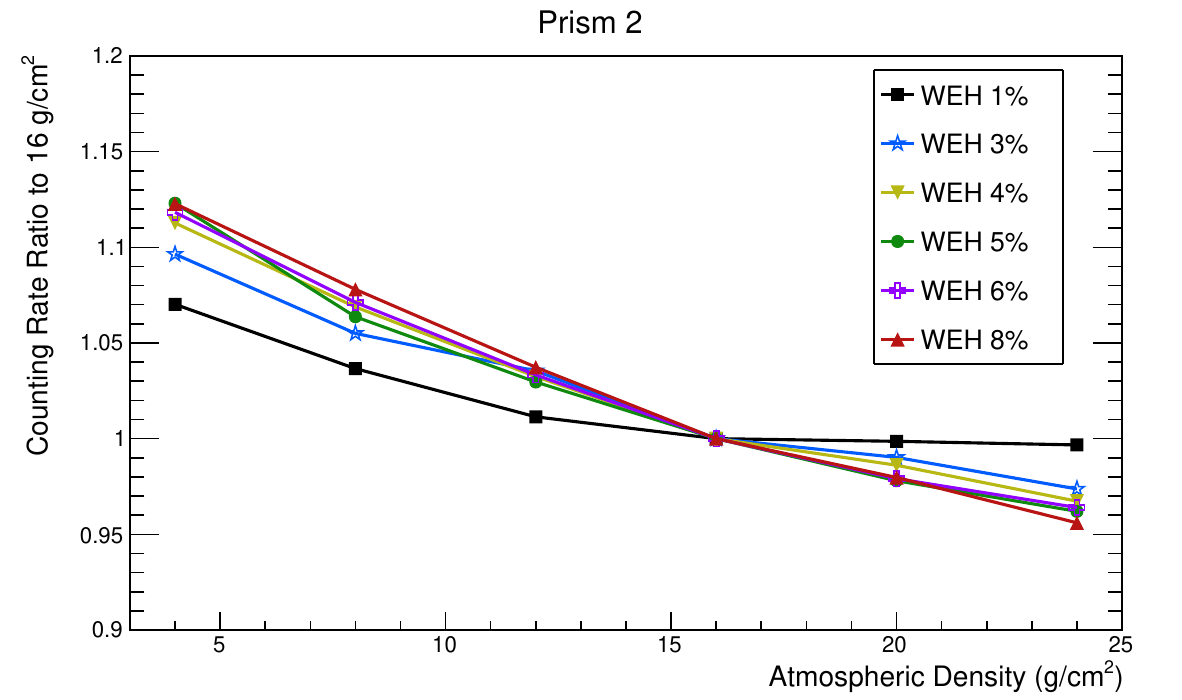}
\includegraphics[width=0.48\textwidth]{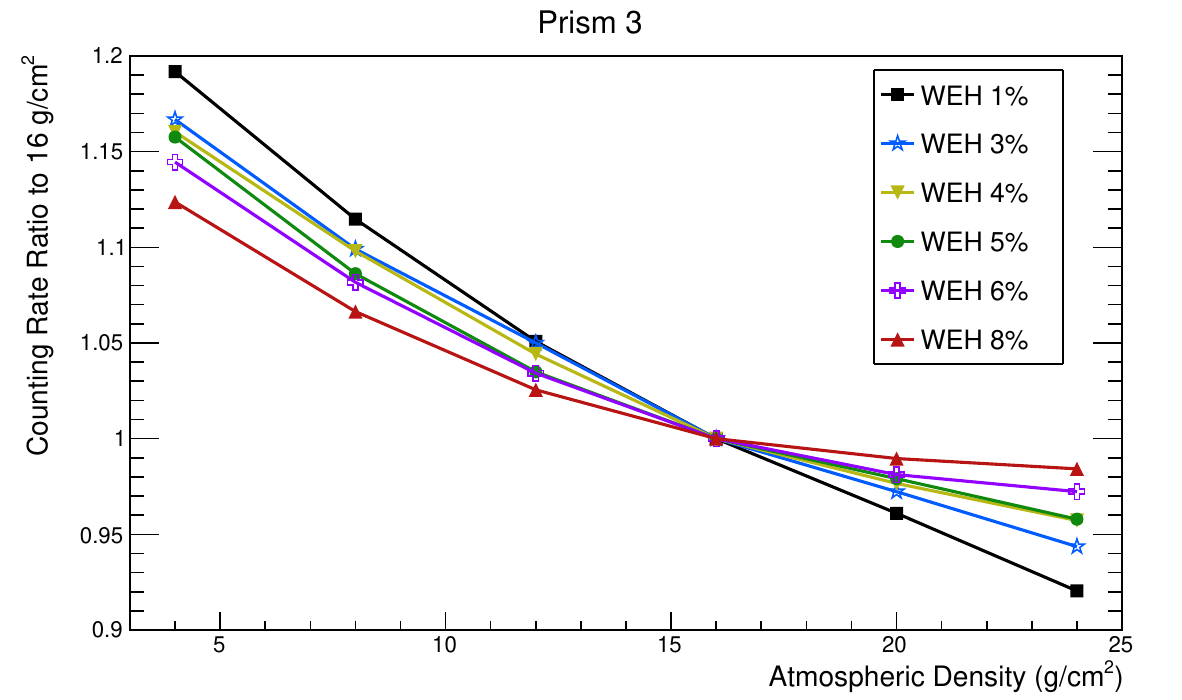}
\includegraphics[width=0.48\textwidth]{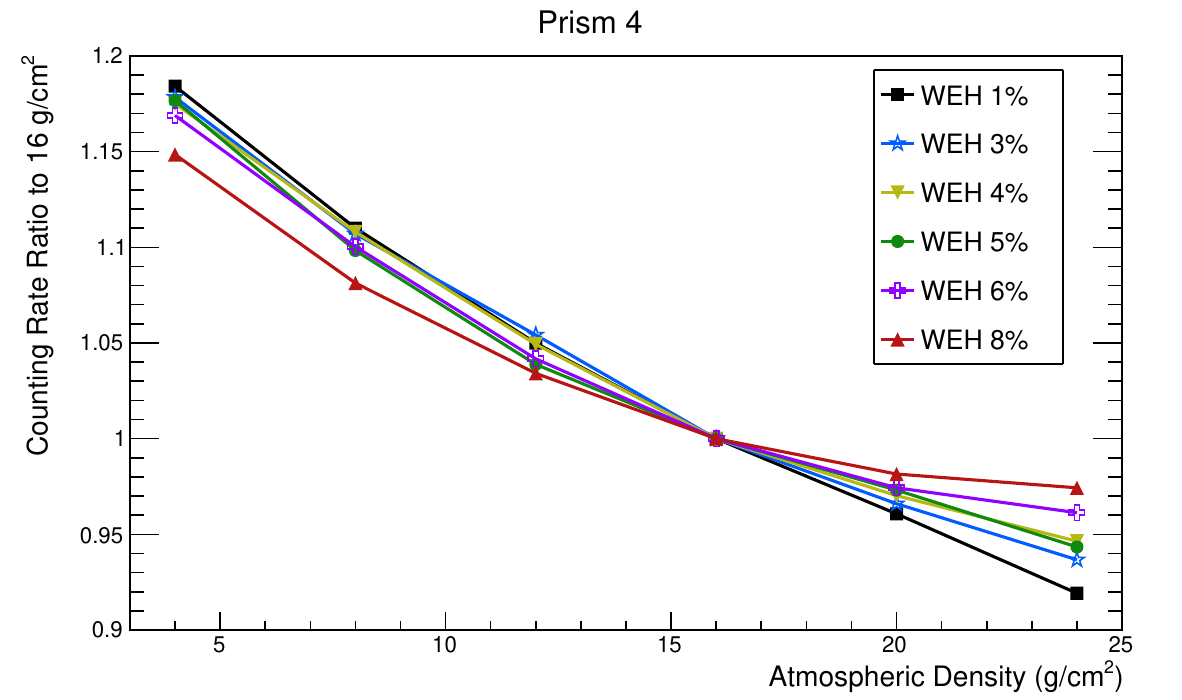}
\caption{Simulated Prism counting rates normalized to an atmospheric density of 16~g/cm$^2$ for different WEH abundances.}
\label{fig:atmosim}
\end{figure}

The resulting ratio of count rates for Prism 1, Prism 2, Prism 3, and Prism 4 as a function of atmospheric density ($\rho_a$) for the different WEH values is shown in Fig.~\ref{fig:atmosim}.  For low WEH content, the correction factor is as much as 19\% for the lowest atmospheric density, however, the typical correction is much smaller. These curves were fit to second-order polynomials, leading to the fit parameters given in Table~\ref{table:fit_params}.

\begin{table}[h]
\centering
\caption{Fit parameters (squared term, linear term, and constant term) for determining atmospheric scaling factor relative to 16~g/cm$^2$.}
\label{table:fit_params}
\begin{tabular}{lcccccc}
Prism/Term & 1\% WEH & 3\% WEH & 4\% WEH & 5\% WEH & 6\% WEH & 8\% WEH \\
\hline
Prism 1 Sqr & 3.169e-4 & 3.133e-4 & 3.752e-4 & 4.160e-4 & 4.265e-4 & 4.237e-4 \\
Prism 1 Lin & -2.227e-2 & -1.990e-2 & -2.077e-2 & -2.143e-2 & -2.060e-2 & -1.880e-2\\
Prism 1 Cns & 1.275 & 1.241 & 1.239 & 1.235 & 1.221 & 1.192 \\
\hline
Prism 2 Sqr &2.831e-4 & 1.827e-4 & 2.405e-4 & 2.956e-4 & 2.545e-4 & 2.088e-4 \\
Prism 2 Lin &-1.145e-2 & -1.114e-2 & -1.393e-2 & -1.608e-2 & -1.485e-2 & -1.418e-2 \\
Prism 2 Cns &1.110 & 1.137 & 1.165 & 1.179 & 1.174 & 1.177 \\
\hline
Prism 3 Sqr &3.148e-4 & 3.132e-4 & 3.751e04 & 4.158e-4 & 4.289e-4 & 4.258e-4 \\
Prism 3 Lin &-2.216e-2 & -1.981e-2 & -2.067e-2 & -2.132e-2 & -2.056e-2 & -1.873e-2 \\
Prism 3 Cts &1.274 & 1.240 & 1.238 & 1.234 & 1.219 & 1.190 \\
\hline
Prism 4 Sqr &2.753e-4 & 3.189e-4 & 3.701e-4 & 4.173e-4 & 4.571e-4 & 4.629e-4 \\
Prism 4 Lin &-2.073e-2 & -2.097e-2 & -2.183e-2 & -2.298e-2 & -2.321e-2 & -2.157e-2 \\
Prism 4 Cts &1.261 & 1.257 & 1.257 & 1.259 & 1.255 & 1.226 \\
\hline
\end{tabular}
\end{table}

To correct out the changes in neutron counting rates due to seasonal variations in the atmosphere, the data were binned for each year in $L_s$.  At each $L_s$ point, the MCD GCM \cite{Forget1999} was used to determine $\rho_a$ for each 2$^{\circ}$$\times$2$^{\circ}$ latitude and longitude bin within the belly band region.  The correction factor based on this $\rho_a$ was then calculated for each of the six simulated WEH values using the fitted parameters.  The published map of WEH derived from the Mars Odyssey high-purity germanium gamma-ray spectrometer \cite{Boynton2007,GRSmap} was then used to determine the WEH of the soil within each bin, and the final correction factor determined by an interpolation of the correction factors covering the range of WEH values.  Once the neutron counting rates were normalized to 16~g/cm$^2$, the GCR correction factor was determined.

\newpage

\bibliographystyle{elsarticle-num}
\bibliography{biblio}

\end{document}